\documentclass[
  journal=pasa,
  manuscript=research-paper,
  year=2020,
  volume=37,
]{cup-journal}

\usepackage{microtype,siunitx,booktabs}
\usepackage{makecell}
\usepackage{hyperref} 
\hypersetup{colorlinks,citecolor=blue,linkcolor=blue,urlcolor=blue}
\usepackage[nolist]{acronym}
\usepackage[version=3]{mhchem}
\usepackage{mathtools}
\usepackage{bm}

\usepackage{adjustbox}
\usepackage[normalem]{ulem}
\usepackage{xcolor}
\newcommand{\snr}{MC\,SNR\,J0519--6902}

\newcommand{\ergcm}[1]{erg\,cm$^{-2}$\,s$^{-1}$}
\newcommand{\cxo}{{\it Chandra}}
\def\xmm{\textit{XMM-Newton}}
\def\ein{\textit{Einstein}}
\def\ergcm[1]{erg\,cm$^{-2}$\,s$^{-1}$}
\def\HI{\hbox{H\,{\sc i}}}
\def\HII{\hbox{H\,{\sc ii}}}

\newcommand{\Halpha}{H${\alpha}$}

\def\HI{\hbox{H\,{\sc i}}}
\def\HII{\hbox{H\,{\sc ii}}}

\def\Halpha{H${\alpha}$}

\def\ergcm{\hbox{erg cm$^{-2}$ s$^{-1}$ }}

\usepackage{xcolor}

\begin{acronym}[AWGN]
\acro{2MASS}{The Two Micron All-Sky Survey}
\acro{30Dor}[30 Dor]{30~Doradus}
\acro{AGN}{active galactic nucleus}
\acrodefplural{AGN}{Active Galactic Nuclei}
\acro{ATCA}{Australia Telescope Compact Array}
\acro{ATNF}{Australia Telescope National Facility}
\acro{ATOA}{Australia Telescope Online Archive}
\acro{AT20G}{Australia Telescope 20 GHz Survey}
\acro{ASKAP}{Australian Square Kilometre Array Pathfinder}
\acro{BL}[BL-Lac]{BL Lacertae Objects}
\acro{CASS}{CSIRO Astronomy and Space Science}
\acro{CABB}{Compact Array Broadband Back-end}
\acro{CHII}[{\sc CHii}]{Compact \textsc{Hii}}
\acro{CSIRO}{Australian Commonwealth Scientific and Industrial Research Organisation}
\acro{CSS}{Compact Steep Spectrum}
\acro{CME}{Coronal Mass Ejection}
\acro{Dec}{Declination}
\acro{DD}{double-degenerate}
\acro{DS9}[\textsc{DS9}]{\textsc{SAOImage DS9}}
\acro{DSS}{Digital Sky Survey}
\acro{ED}[$n_e$]{electron density}
\acro{EMU}{Evolutionary Map of the Universe}
\acro{ev}[eV]{electronvolt\acroextra{: 1 eV $\approx 1.6 \times 10^{-19}$ J}}
\acro{FITS}[\textsc{Fits}]{Flexible Image Transport System}
\acro{FRB}{Fast Radio Bursts}
\acro{FRII}{Fanaroff-Riley II}
\acrodefplural{FRII}{Fanaroff-Riley II's}
\acro{FRI}{Fanaroff-Riley I}
\acrodefplural{FRI}{Fanaroff-Riley I's}
\acro{FSRQ}{Flat Spectrum Radio Quasars}
\acro{FWHM}{Full Width at Half-Maximum}
\acro{GLEAM}{GaLactic Extragalactic All-sky MWA}
\acro{GMRT}{Giant Meterwave Radio Telescope}
\acro{GRG}{Giant Radio Galaxy}
\acro{GPS}{Gigahertz Peak Spectrum}
\acro{HEMT}{High Electron Mobility Transistor}
\acro{HFP}[HFP]{High Frequency Peaker}
\acro{HzRGs}{High Redshift Radio Galaxies}
\acro{HVS}{hypervelocity stars}
\acro{pHFP}[pHFP]{Potential High Frequency Peaker}
\acro{HST}{\textit{Hubble Space Telescope}}
\acro{IAU}{International Astronomical Union}
\acro{IFRSs}{Infrared Faint Radio Sources}
\acro{ISM}{Interstellar Medium}
\acro{IR}{Infrared}
\acro{JY}[Jy]{Jansky\acroextra{, 1 Jy = $10^{-26} \times \mathrm{W~ m}^{-2}~\mathrm{Hz}^{-1}$}} 
\acro{LLS}{Largest Linear Size}
\acro{LFAA}{Low-Frequency Aperture Array}
\acro{LMC}{Large Magellanic Cloud}
\acro{LG}{Local Group}
\acro{LSO}[LSOs]{Large Scale Objects}
\acro{MACHO}{Massive Astrophysical Compact Halo Objects}
\acro{MC}[MCs]{Magellanic Clouds}
\acro{mc}[MC]{Magellanic Cloud}    %singular 
\acro{MCELS}{Magellanic Cloud Emission Line Survey}
\acro{MW}{Milky Way}
\acro{MIRIAD}[\textsc{Miriad}]{Multichannel Image Reconstruction, Image Analysis and Display}
\acro{MIT}{Massachusetts Institute of Technology}
\acro{MOST}{Molonglo Observatory Synthesis Telescope}
\acro{MQS}{Magellanic Quasars Survey}
\acro{MRC}{Molonglo Reference Catalogue of Radio Sources}
\acro{MWA}{Murchison Widefield Array}
\acro{NAT}{Narrow-Angle Tail}
\acro{NRAO}{National Radio Astronomy Observatory}
\acro{NVSS}{NRAO VLA Sky Survey}
\acro{ORCs}{Odd Radio Circles}
\acro{ORC}{Odd Radio Circle}
\acro{OPAL}{Online Proposal Applications \& Links}
\acro{OVV}{Optically Violent Variable Quasars}            
\acro{PAF}{Phased Array Feed}
\acro{pc}{parsec\acroextra{: 1 pc $\simeq 3.09 \times 10^{16}$ m}}
\acro{PMN}{Parkes-MIT-NRAO}
\acro{PNe}{Planetary Nebulae}
\acro{PN}{Planetary Nebula}
\acro{PWN}{Pulsar Wind Nebula}
\acro{QSO}{Quasi-Stellar Object}
\acro{RA}{Right Ascension}
\acro{RFI}{Radio-Frequency Interference}
\acro{RMS}[rms]{Root Mean Squared}
\acro{RM}[$RM$]{rotation measure}
\acro{SDSS}{Sloan Digital Sky Survey}
\acro{SED}{spectral energy distribution}
\acro{SI}[$\alpha$]{Spectral Index\acroextra{, $S \propto \nu^\alpha$}}
\acro{SKA}{Square Kilometre Array}
\acro{SMASH}{Survey of the MAgellanic Stellar History}
\acro{SMB}{Super Massive Blackholes}
\acro{SMC}{Small Magellanic Cloud}
\acro{SN}{Supernova}
\acro{SNe}{Supernovae}
\acro{SNR}{Supernova Remnant}
\acrodefplural{SNRs}{Supernova Remnants}
\acro{SD}{single-degenerate}
\acro{SUMSS}{Sydney University Molonglo Sky Survey}
\acro{SMBH}{Super Massive Black Hole}
\acro{TOPCAT}[\textsc{Topcat}]{Tool for OPerations on Catalogues And Tables}
\acro{USS}{Ultra Steep Spectrum}
\acro{WBAC}{Wide-Band Analogue Correlator}
\acro{WIFES}[WiFeS]{Wide-Field Spectrograph}
\acro{WISE}{Wide-Field Infrared Survey Explorer}
\acro{VLA}{Very Large Array} 
\acro{VLBI}{Very Long Baseline Interferometry} 
\acro{VLSR}[\textbf{$v_{lsr}$}]{Velocity in the Line of Sight}
\acro{WD}{white dwarf}
\end{acronym}

\title{A New Radio Continuum Study of the Large Magellanic Cloud Supernova Remnant \snr}

\author{Rami Z. E. Alsaberi}
\affiliation{Faculty of Engineering, Gifu University, 1-1 Yanagido, Gifu 501-1193, Japan}
\alsoaffiliation{Western Sydney University, Locked Bag 1797, Penrith NSW 2751, Australia}
\email[R. Z. E. Alsaberi]{ramy\_z@yahoo.com}

\author{Miroslav D. Filipovi\'c}
\affiliation{Western Sydney University, Locked Bag 1797, Penrith NSW 2751, Australia}
\author{Hidetoshi Sano}
\affiliation{Faculty of Engineering, Gifu University, 1-1 Yanagido, Gifu 501-1193, Japan}
\alsoaffiliation{Center for Space Research and Utilization Promotion (c-SRUP), Gifu University, 1-1 Yanagido, Gifu 501-1193, Japan}
\alsoaffiliation{National Astronomical Observatory of Japan, Mitaka, Tokyo 181-8588, Japan}

\author{Shi Dai}
\affiliation{Western Sydney University, Locked Bag 1797, Penrith NSW 2751, Australia}

\author{Frank Haberl}
\affiliation{Max-Planck-Institut f\"{u}r extraterrestrische Physik, Gie{\ss}enbachstra{\ss}e 1, D-85748 Garching, Germany}

\author{Patrick J. Kavanagh}
\affiliation{Department of Physics, Maynooth University, Maynooth, Co Kildare, Ireland}

\author{Denis Leahy}
\affiliation{Department of Physics and Astronomy, University of Calgary, University of Calgary, Calgary, Alberta, T2N 1N4, Canada}

\author{Pierre Maggi}
\affiliation{Observatoire Astronomique de Strasbourg, Universit\'e de Strasbourg, CNRS, 11 rue de l'Universit\'e, F-67000 Strasbourg, France}

\author{Gavin Rowell}
\affiliation{School of Physical Sciences, The University of Adelaide, Adelaide 5005, Australia}

\author{Manami Sasaki}
\affiliation{Remeis Observatory and ECAP, Universit\"{a}t Erlangen-N\"{u}rnberg, Sternwartstra{\ss}e 7, D-96049 Bamberg, Germany}

\author{Ivo R. Seitenzahl}
\affiliation{Heidelberg Institute for Theoretical Studies (HITS), Schloss-Wolfsbrunnen Weg 35, 69118, Heidelberg, Germany}

\author{Dejan Uro\v sevi\' c}
\affiliation{Department of Astronomy, Faculty of Mathematics, University of Belgrade, Studentski trg 16, 11000 Belgrade, Serbia} 
\alsoaffiliation{Isaac Newton Institute of Chile, Yugoslavia Branch}

\author{Jeffrey Payne}
\affiliation{Western Sydney University, Locked Bag 1797, Penrith NSW 2751, Australia}

\author{Zachary J. Smeaton}
\affiliation{Western Sydney University, Locked Bag 1797, Penrith NSW 2751, Australia}

\author{Sanja Lazarevi\'c}
\affiliation{Western Sydney University, Locked Bag 1797, Penrith NSW 2751, Australia}
\alsoaffiliation{CSIRO Space and Astronomy, Australia Telescope National Facility, PO Box 76, Epping, NSW 1710, Australia}
\alsoaffiliation{Astronomical Observatory, Volgina 7, 11060 Belgrade, Serbia}
\doi{}

\received {dd Mmm YYYY}
\revised  {dd Mmm YYYY}
\accepted {dd Mmm YYYY}
\published{dd Mmm YYYY}


\begin{document}

%%%%%%%%%% CONTENT %%%%%%%%%%
% \clearpage
% \begingroup
% \let\clearpage\relax
% \tableofcontents
% \endgroup 

% \vspace{15mm}
% %TO-DO LIST: \\
% %-- address all comments

% \newpage
%%%%%%%%%% CONTENT END %%%%%%%%%%

%%%%%%%%%% ABSTRACT %%%%%%%%%%

\begin{abstract}
We present a new radio continuum study of the Large Magellanic Cloud supernova remnant (SNR) \snr. With a diameter of $\sim$8\,pc, this SNR shows a radio ring-like morphology with three bright regions toward the north, east, and south. Its linear polarisation is prominent with average values of $5\pm1$\% and $6\pm1$\% at 5500 and 9000\,MHz, and we find a spectral index of ${-0.62\pm0.02}$, typical of a young SNR. The average  rotation measure is estimated at ${-124\pm83}$\,rad\,m$^{-2}$ and the magnetic field strength  at ${\sim11}\,\mu$G. We also estimate an equipartition magnetic field of ${72\pm 5}\,\mu$G and  minimum explosion energy of E$_{\rm min}$~=~2.6$\times10^{48}$\,erg. Finally, we identified an \HI\ cloud that may be associated with \snr, located in the southeastern part of the remnant, along with a potential wind-bubble cavity. %

%Because of the remnant's asymmetrical morphology \textbf{with respect to brightness} and the possible existence of a wind-bubble cavity, we favour a supernova Type\,Ia single-degenerate scenario.  
\end{abstract}
%%%%%%%%%% ABSTRACT END %%%%%%%%%%

%%%%%%%%%% 1 INTRODUCTION %%%%%%%%%%
\acresetall
\section{Introduction}
\label{sec:intro}
\acp{SNR} play an essential role in the structure of galaxies, enriching the \ac{ISM} as well as having a significant impact on the structure and physical properties of the \ac{ISM}. The study of \acp{SNR} in our own Galaxy is not ideal because  %of the various stages of dense Galactic Plane 
of difficulties of making accurate distance measurements. Instead, we study \acp{SNR} in nearby, small, dwarf galaxies, including the \ac{LMC} located at a distance of 50\,kpc \citep{mt,2019Natur.567..200P}. Because all objects within this galaxy are located at approximately the same distance, physical measurements, including the physical size, are more reliable. The almost face-on orientation of the  \ac{LMC} \citep[inclination angle of $\sim$35$^{\circ}$,][]{2001AJ....122.1807V} allows deep, high-resolution (spatial and spectral) multi-frequency observations \citep{2016A&A...585A.162M,2017ApJS..230....2B}. Moreover, the \ac{LMC} contains active star-forming regions, and is located away from %DL is located outside of 
the Galactic plane where absorption by gas and dust is reasonably low.
%\cite{1994PASAu..11...99D} stated that this \ac{SNR} is similar to Tycho or Keplers \ac{SNR} in the Milky Way.

%\cite{1995A&AS..111..311F} measured an integrated flux density measurement of 57\,mJy at 3\,cm. \cite{1998A&AS..127..119F} re observed this \ac{SNR} with the Parkes radio-telescope at 6\,cm, estimating an integrated flux density of 72\,mJy.   

To date, numerous studies have investigated \acp{SNR} within the \ac{LMC}. \cite{2016A&A...585A.162M} reported 59 \acp{SNR} as X-ray emitters, and the radio-continuum studies of \citet{2017ApJS..230....2B} have added 15 candidates to that list. \cite{2017ApJ...837...36L} studied the most energetic and brightest \ac{LMC} \acp{SNR} using an analysis of the \cite{2016A&A...585A.162M} \ac{SNR} sample. \citet{2019MNRAS.490.5494M,2021MNRAS.504..326M} promoted two additional objects as bona fide \acp{SNR}; \cite{2021MNRAS.500.2336Y} discovered two new optical \acp{SNR} and 14 candidates. Most recently, \cite{2022MNRAS.515.4099K,2023MNRAS.518.2574B,2022MNRAS.512..265F,2024A&A...692A.237Z} confirmed eight new \ac{LMC} \acp{SNR} and suggested an additional 15 as `good' candidates. This currently gives us a total of 71 confirmed \ac{LMC} \acp{SNR} and an additional 20 candidates. Thirteen \acp{SNR} are confirmed and two candidates are possible Type\,Ia remnants  \citep{2017ApJS..230....2B}.

%{\bf You would need here a paragraph here that will talk generally about MCs SNRs and then specifically about the population of Type Ia's. Use/mention studies/papers by Bozzetto+19, Maggi+XX, Yew+21...}

\snr\ (also known as LHG\,26) was initially discovered using the \ein\ observatory \citep{1981ApJ...248..925L}. \cite{1982ApJ...261..473T} confirmed  the \ac{SNR} is Balmer-dominated with a broad H$\alpha$ component. They reported an X-ray angular size of $\sim$30$''$ and an optical angular size of 28$''$ with a forward shock velocity of $2900\pm400$\,km\,s$^{-1}$. They suggest  the remnant is expanding into a low-density region composed of neutral hydrogen, inferring a Type\,Ia \ac{SN} with an estimated progenitor mass of between 1.2 and 4.0\,M$_{\odot}$. \cite{2010A&A...519A..11K} used \cxo\ and \xmm\ observations to estimate the velocity for the forward shock of $2770\pm500$\,km\,s$^{-1}$ with no strong non-thermal X-ray continuum emission.  
%DL grammar correction
\cite{2018ApJ...862..148H} used optical data to report a forward shock velocity of 2650\,km\,s$^{-1}$.

\cite{1988AJ.....96.1874C} associated \snr\ with the nearby (200\,pc) population\,II OB association -- LH41. \cite{2010AJ....140..584D} found no young stellar object associated with this \ac{SNR}, while \cite{2012ApJ...747L..19E} reported that this \ac{SNR} could have only been a Type\,Ia \ac{SN} event resulting from a supersoft X-ray source or a double degenerate system. 

%\cite{1999A&AS..139..277H} observed this \ac{SNR} with the ROSAT and gave the association [HP] 789
\cite{1983ApJS...51..345M} estimated a radio spectral index \footnote{This is defined by $S_{\nu}$~$\propto$~$\nu^\alpha$, where $S_{\nu}$ is flux density, $\nu$ is frequency, and $\alpha$ is the spectral index.} for \snr\ of $\alpha$=--0.6, while  \cite{1984AuJPh..37..321M} reported an index of --0.65. \cite{2012SerAJ.185...25B} listed a spectral index of $-0.53\pm0.07$, which still suggested predominantly synchrotron  emission (e.g. \citealt{2021pma..book.....F}) from a typical young \ac{SNR} \citep{2012SSRv..166..231R,2014SerAJ.189...15G,2017ApJS..230....2B,2019A&A...631A.127M}.

\cite{1982ApJ...261..473T} inferred an age for \snr\ of $\sim$500\,yrs while \cite{1991ApJ...375..652S} reported an age between 500 -- 1500\,yrs. \cite{2005Natur.438.1132R} used a light echo method to estimate an age of $600\pm200$\,yrs. Based on \cxo\ and \xmm\ observations, \cite{2010A&A...519A..11K} suggested an age of $450\pm200$\,yrs.  \cite{2017ApJ...837...36L} used X-ray emission and temperature to derive an age of $\sim$2700\,yrs, assuming a shock radius of 4.1\,pc expanding in a uniform \ac{ISM}. 
%\Denis{The following should be moved to section 4, because it is new}

\cite{1995AJ....109..200D} observed this remnant with the \ac{ATCA} at 1472 and 2368\,MHz. They estimated an average fractional polarisation across the remnant of $1.5\pm0.6$\% and $4.1\pm0.6$\%, respectively. \cite{2012SerAJ.185...25B}, who estimated the diameter of this \ac{SNR} at $\sim$8\,pc, found average fractional polarisation values of $\sim$2.2\% and $\sim$3.2\% at 5500 and 9000\,MHz, respectively. They also calculated a \ac{RM} for the entire remnant of $\sim$10\,rad\,m$^{-2}$. 

%They broke down the remnant into three regions: field 01 (north-east), field 02 (north-west), and field 03 (south). The mean $RM$ value toward field 01 was --272\,rad\,m$^{-2}$, 462\,rad\,m$^{-2}$ toward field 02, and 697\,rad\,m$^{-2}$ toward field 03.

\cite{2007RMxAA..43...33V} estimated the magnetic field of this \ac{SNR}, using both a classic and revised equipartition formula  to obtain results of 186 and 270\,$\mu$G, respectively. \cite{2012SerAJ.185...25B} used the modified equipartition model from \cite{2012ApJ...746...79A} to estimate the magnetic field to be $\sim$171\,$\mu$G with a minimum energy of E$_{\rm min}$~=~1.8$\times10^{49}$\,erg. 

%{\color{red}Jeff: You may want to clarify "classic" and "revised" formulas; are you talking about Arbutina and Urosevic? Also which is "modified" formula, do you mean Urosevic's 2018 model?}

\cite{2015MNRAS.449.1441K} reported a method based on state-of-the-art 3D simulations of thermonuclear SN explosions, coupled with hydrodynamic calculations of \ac{SNR} evolution, while making use of the most up-to-date atomic data to suggest \snr\ originated from an oxygen-rich merger. \cite{2019PhRvL.123d1101S} discovered  optical ([Fe\,XIV] 5303\,\Angstrom) emission associated with this \ac{SNR}. \cite{2019ApJ...886...99L} searched for a surviving companion of \snr\ and found a candidate run-away companion star moving at a radial velocity of $182$\,km\,s$^{-1}$. %\textbf{0 must be a typo to fix}%DL.

In this paper, we present new high-resolution \ac{ATCA} radio continuum images of \snr\ at 5500 and 9000\,MHz. In Section~\ref{obs} we describe our observations and data analyses. Our new findings and discussion are listed in Section~\ref{res}. Finally, our conclusions are outlined in Section~\ref{con}. 

%%%%%%%%%% INTRODUCTION END %%%%%%%%%%

%%%%%%%%%% 2 OBSERVATIONS AND DATA REDUCTION %%%%%%%%%%

%

\section{Observation and Data Analysis}
\label{obs}

\subsection{Radio Continuum Observations}
\label{radio_obs}

We observed \snr\ with \ac{ATCA} (project codes: CX454 and CX310). \ac{ATCA} archival\footnote{\ac{ATOA}, hosted by the \ac{ATNF}: \url{https://atoa.atnf.csiro.au}} data (project code: C634) were also used to produce the high-resolution and sensitive images (see Table~\ref{tab1}). All observations were carried out in ``snap-shot'' mode, with 1-hour of integration over a 12-hour minimum using the \ac{CABB} (2048\,MHz bandwidth) at wavelengths of 3/6\,cm ($\nu$~=~4500--6500 and 8000--10000\,MHz centred at 5500 and 9000\,MHz) totalling $\sim$580\,minutes integration. The primary (flux density) calibrator, PKS\,B1934--638, with a flux density of 4.96\,Jy for 5500\,MHz and 2.70\,Jy for 9000\,MHz, and the secondary (phase) calibrator, PKS\,B0530--727, with a flux density of 0.77\,Jy for 5500\,MHz and 0.82\,Jy for 9000\,MHz were used for all three observing days.

\textsc{miriad}\footnote{\url{http://www.atnf.csiro.au/computing/software/miriad/}} \citep{1995ASPC...77..433S} and \textsc{karma}\footnote{\url{http://www.atnf.csiro.au/computing/software/karma/}} \citep{1995ASPC...77..144G} software packages were used for data reduction and analysis. Imaging was completed using the multi-frequency synthesis \textsc{invert} task with natural Briggs weighting (robust=0 for both 5500 and 9000\,MHz). Beam sizes included $2.3''\times1.7''$ and $1.4''\times1.0''$ for 5500 and 9000\,MHz images, respectively. \textsc{mfclean} and \textsc{restor} algorithms allowed the images to be deconvolved, with primary beam correction applied using  \textsc{linmos}. We followed the same process for stokes Q and U images using a smoothed resolution of $5''\times5''$ (see Section~\ref{pol}).

%%%%%%%%%%%%%%%%%%%%%%%%%%%%%%%%%%%%% Table 1 %%%%%%%%%%%%%%%%%%%%%%%%%%%%%%%%%%%%%%
%\begin{table*}
%\caption{\ac{ATCA} observations of \snr.}
%\centering
%\scriptsize
%\begin{tabular}{@{}lllccccccl@{}}
%\hline\hline
%Date                     &Project & Array          & Channels & Bandwidth &  Frequency  &Primary        & Secondary       & Integrated Time& References \\
%                         &  Code  & Configuration  &           &    (GHz)  &$\nu$ (MHz) &Calibrator  &Calibrator    &   (minutes)&
%            \\
% \hline%
%15$^\mathrm{th}$~Nov~2011&  C634  &    EW367       &  2049 &    2.048         & 5500 \& 9000        &PKS\,B1934--638 & PKS\,B0530--727 & $\sim$30 & \cite{2012SerAJ.185...25B}\\
%01$^\mathrm{st}$~Jan~2015&  CX310 &    6A          &  2049 &    2.048         & 5500 \& 9000        &PKS\,B1934--638 & PKS\,B0530--727 & $\sim$50& Our observation\\
%28$^\mathrm{th}$~Dec~2019&  CX454 &   1.5C         &  2049 &    2.048         & 5500 \& 9000        &PKS\,B1934--638 & PKS\,B0530--727 & $\sim$500& Our observation\\
%\hline\hline
%\end{tabular}
%\label{tab1}
%\medskip
%\end{table*}
%%%%%%%%%%%%%%%%%%%%%%%%%%%%%%%%%%%%%%%%%%%%%%%%%%%%%%%%%%%%%%%%%%%

%%%%%%%%%%%%%%%%%%%%%%%%%%%%%%%%%%%%% Table 1 %%%%%%%%%%%%%%%%%%%%%%%%%%%%%%%%%%%%%%
\begin{table}
\caption{\ac{ATCA}  observations of \snr.}
\centering
\scriptsize
\begin{tabular}{@{}lcccl@{}}
\hline\hline
Date                     & Project& Array          & Int. Time on source & References \\
                         &  Code  & Config.  &   (minutes)     &
            \\
 \hline%
15$^\mathrm{th}$~Nov~2011&  C634  &    EW367       &  $\sim$30      & \cite{2012SerAJ.185...25B}\\
01$^\mathrm{st}$~Jan~2015&  CX310 &    6A          &  $\sim$50      & This work\\
28$^\mathrm{th}$~Dec~2019&  CX454 &   1.5C         &  $\sim$500     & This work\\
\hline\hline
\end{tabular}
\label{tab1}
\medskip
\end{table}
%%%%%%%%%%%%%%%%%%%%%%%%%%%%%%%%%%%%%%%%%%%%%%%%%%%%%%%%%%%%%%%%%%%

\subsection{\texorpdfstring{\HI}{H} Observations}
\label{h_obs}

Archival \HI\ data \citep{2003ApJS..148..473K} were obtained using the \ac{ATCA} and  Parkes 64-m telescope. The angular resolution of the data is $60''$, corresponding to a spatial resolution of $\sim$15\,pc at the distance of the \ac{LMC}. The typical noise level is $\sim$2.4~K at a velocity resolution of 1.689\,km\,s$^{-1}$.

\subsection{Chandra Observations} \label{chandra}

We used archival X-ray data obtained using \cxo\, for which the observation IDs (Obs IDs) are 118 (PI: S. Holt), 11241, 12062, and 12063 (PI: J. P. Hughes);  the data have been published by several authors \citep[e.g.,][]{2010A&A...519A..11K, 2012ApJ...747L..19E,2016AJ....151..161S}. All datasets were taken using the Advanced CCD Imaging Spectrometer S-array (ACIS-S3) in June 2000 (Obs ID 118),  December~2009 (Obs IDs 11241 and 12062), and in February~2010 (Obs ID 12063). We utilised  Chandra Interactive Analysis of Observations \citep[CIAO,][]{2006SPIE.6270E..1VF} version 4.12 with CALDB 4.9.1 for data reprocessing and imaging. ``Chandra\_repro'' and ``merge\_obs'' scripts created an exposure-corrected, energy-filtered image  at 0.5--7.0\,keV (hereafter referred to as ``broadband'') with a total effective exposure of $\sim$91\,ks. Lastly, we smoothed the image with a Gaussian kernel of 1$''$, \ac{FWHM}.

\subsection{Optical HST Observations}  %don't use \ac{} commands within section statements, thanks jlp
 \label{hst}

The \ac{HST} image of \snr\ was downloaded from the Mikulski Archive for Space Telescopes\footnote{\url{https://mast.stsci.edu/portal/Mashup/Clients/Mast/Portal.html}} portal. All details regarding these \ac{HST} observations and their data reduction are described in \cite{2012ApJ...747L..19E}.

\section{Results and Discussion}
\label{res}

\subsection{Radio Morphology}
\label{radio_mor}

\begin{figure*}[ht!]
\begin{center}
\includegraphics[width=\textwidth,trim=0 0 0 0,clip]{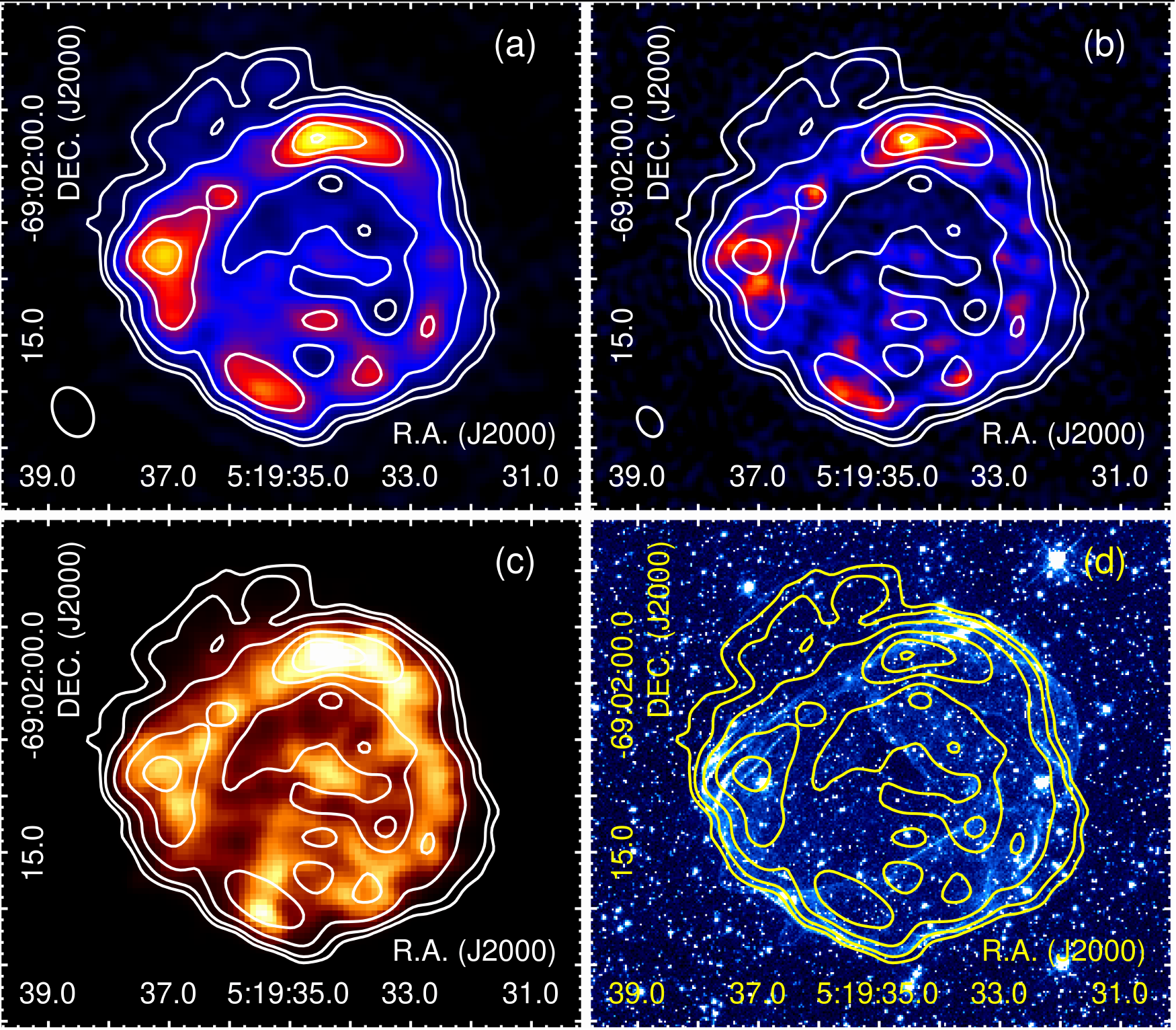}\\
\caption{\emph{(a)} \ac{ATCA} image at 5500\,MHz of \snr. The ellipse in the lower left corner represents a synthesised beam of $2.3''\times1.7''$ with a p.a. of 62$^\circ$. \emph{(b)} \ac{ATCA} image at 9000\,MHz. The ellipse in the lower left corner represents a synthesised beam of $1.4''\times1.0''$ with a p.a. of 61$^\circ$. The rms noise of the 5500 and 9000\,MHz images is 20\,$\mu$Jy\,beam$^{-1}$ and 17\,$\mu$Jy\,beam$^{-1}$, respectively. \emph{(c)} \cxo\ broadband image. \emph{(d)} \ac{HST} image \citep[combination of the V-band and \Halpha\ images,][]{2012ApJ...747L..19E}. 5500\,MHz contour lines overlaid on each image are: 0.05, 0.09, 0.2, 0.4, 0.6, and 0.8\,mJy\,beam$^{-1}$.
\label{Fig1}}
\end{center}
\end{figure*}

\snr\ has a ring-like morphology with three bright regions towards the north, \textbf{east}, and south \citep{2012SerAJ.185...25B}. In Figure~\ref{Fig1}, the new \ac{ATCA} images at 5500 and 9000\,MHz are compared with the \cxo\ and \ac{HST} images. \ac{RMS} noise for these high resolution \ac{ATCA} images at 5500 and 9000\,MHz are $\sim$20 and $\sim$17\,$\mu$Jy\,beam$^{-1}$, respectively; one order of magnitude better than that obtained by \citet{2012SerAJ.185...25B}. %We used the \ac{ATCA} sensitivity calculator\footnote{\url{https://www.narrabri.atnf.csiro.au/myatca/interactive_senscalc.html}} to estimate the surface brightness sensitivity for \ac{ATCA} images\footnote{We used robust~=~0 and typical weather for all images. \ac{RMS} noise (sensitivity) of 0.02\,mJy\,beam$^{-1}$ with an array of 1.5\,km for our 5500\,MHz image, and \ac{RMS} noise of 0.3\,mJy\,beam$^{-1}$ with an array of EW367 for 5500\,MHz image from \cite{2012SerAJ.185...25B}. We excluded antenna 6 (CA06: NO) from the image reported by \cite{2012SerAJ.185...25B}.}. Our new 5500\,MHz image achieved a surface brightness sensitivity of $\sim$132\,mK, compared to the $\sim$20\,mK surface brightness sensitivity of the image reported by \cite{2012SerAJ.185...25B}. However, we could not use this calculator to estimate the surface brightness sensitivity for the images from \cite{1995AJ....109..200D} since it was specifically designed for \ac{CABB} data.

The figure clearly shows a radio emission coincident %DL consistent 
with X-ray emission suggesting relativistic electrons are associated with the shock-heated gas. %which is an indication of dominant non-thermal radio emission 
%\Denis{ coincidence of radio and X-ray does not necessary indicate non-thermal radio, rather the radio spectral index does.} 
Interestingly, the optical emission is outside  %outside DL
both the radio continuum and X-ray emission (Figure~\ref{Fig1}),
%DL: I think this is common in SNRs, I don't think it is signature of stellar wind. suggested rewording:
%which could be caused by the stellar wind of the progenitor and thus is indicative of forward shock.
indicating the optical emission is located at the forward shock.
The radio morphology of this \ac{SNR} is asymmetric in terms of variation in brightness. As noted by \cite{2012SerAJ.185...25B}, it features three bright regions located in the north, east, and south 
%moderately asymmetric-moderately is not a precise term because the reader can't quantitativly measure "moderate"
%DL but not strongly asymmetric, most SNRs have similar asymmetry, some a lot more.
(Figure~\ref{Fig1}), resembling the structure of \ac{LMC} \ac{SNR} N\,103B \citep{2019Ap&SS.364..204A} and Kepler in the \ac{MW} \citep{2002ApJ...580..914D}. %
%As suggested by \cite{2019Ap&SS.364..204A}, SN Type\,Ia \ac{SD} systems may likely be asymmetrical. 
Moreover, our new images show a faint structure on the north-east side of \snr\ which was not present in previous images (see Figure~\ref{Fig1}). Although not statistically significant ($\bm{<3\sigma}$), this feature suggests that the \ac{SNR} extends further in the north-east direction compared to the other sides (see text for details).
%\Denis{ because it is not particularly asymmetric I recommend omitting the following. This indicates that \snr\ maybe a product of the \ac{SD} scenario.}

%
%%%%%%%%%%%%%%%%%%%%%%%%%%%%%%%%% Fig 1 %%%%%%%%%%%%%%%%%%%%%%%%%%%%%%%%%%%%%%%%%%%%
%%%%%%%%%%%%%%%%%%%%%%%%%%%%%%%%%%%%%%%%%%%%%%%%%%%%%%%%%%%%%%%%%%%%%%%%%%%%%

%%%%%%%%%%%%%%%%%%%%%%%%%%%%%%%%%%%%%%%%%%%%% Fig. 2 %%%%%%%%%%%%%%%%%%%%%%%%%%%%%%%
% \begin{figure*}[ht!]
% \begin{center}
% %\includegraphics[width=\columnwidth]{fig1.eps}
% \includegraphics[width=\textwidth,trim=0 0 0 0,clip]{figures/RGB_final.eps}\\
% \caption{RGB image of \snr; red represents the \ac{ATCA} image at 5500\,MHz, green represents the \ac{HST} image \citep{2012ApJ...747L..19E}, and blue represents the \textit{Chandra} broadband image.}\label{Fig RGB}
% \end{center}
% \end{figure*}

%%%%%%%%%%%%%%%%%%%%%%%%%%%%%%%%%%%%%%%%%%%%%%%%%%%%%%%%%%%%%%%%%%%%%%%%%%%%%%%%%%%%%%
We used the Minkowski tensor analysis tool BANANA\footnote{\url{https://github.com/ccollischon/banana}} \citep{2021A&A...653A..16C} to determine the centre of expansion. This tool searches for filaments and calculates normal lines and line density maps. 
For a perfect shell, all lines should meet at the centre position inside the \ac{SNR}, where the expansion must have started \citep[see][for more details]{2021A&A...653A..16C}. Since real sources deviate from this ideal picture, we circumvented this problem by smoothing the line density map using a circle with a diameter of 40 pixels. We then took the centre position of the pixel where the line density was the highest. Using the same method, we performed the tensor analysis separately for the 5500\,MHz, 9000\,MHz, and \cxo\ broad-band images. From this, we calculated the mean centre position and the $1\sigma$ uncertainty of the mean.

\begin{figure*}
\centering
\includegraphics[scale=0.9,trim=20 1 60 1.5,clip]{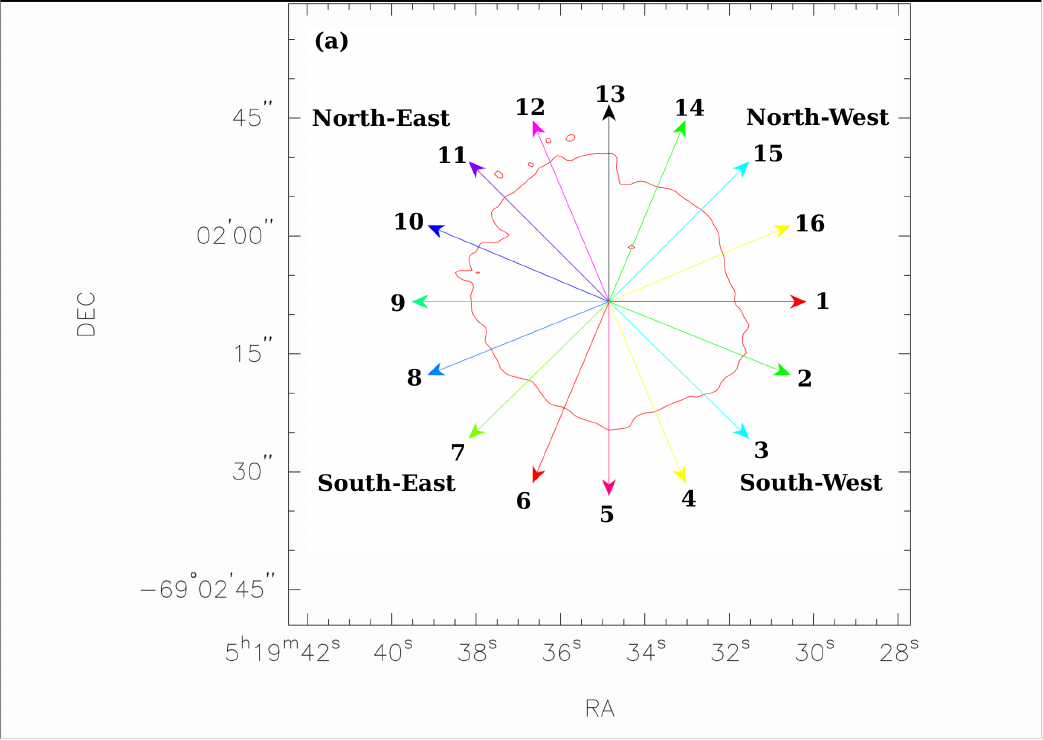}\\
\includegraphics[scale=0.280,trim=10 0 100 50,clip]{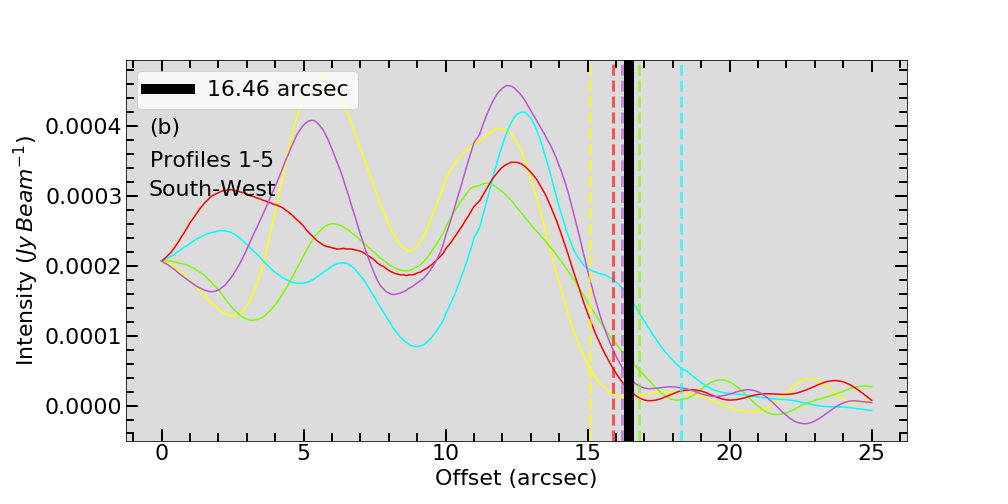}
\includegraphics[scale=0.280,trim=10 0 100 60,clip]{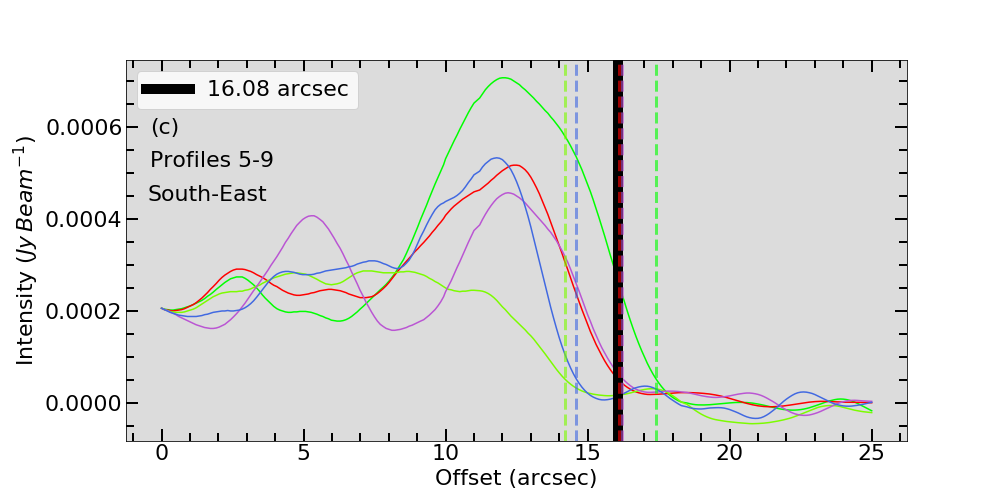}\\
\includegraphics[scale=0.280,trim=10 0 100 60,clip]{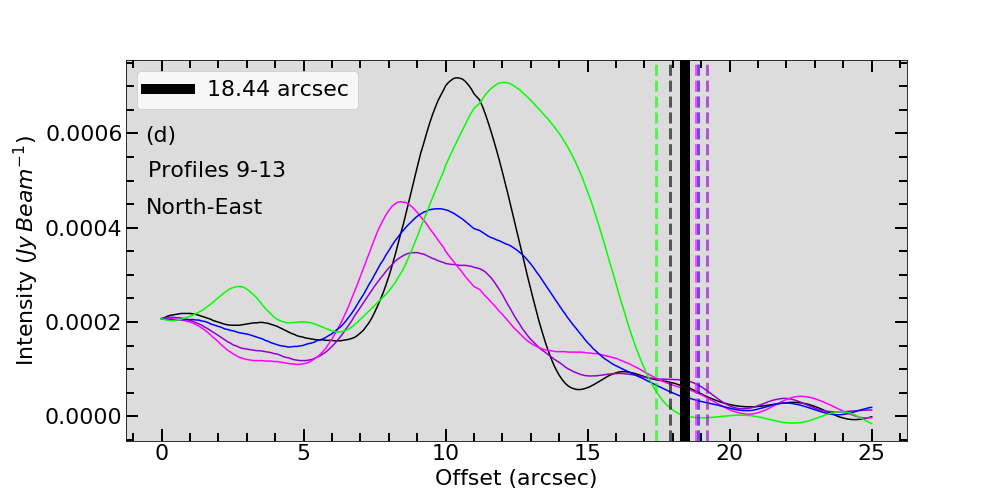}
\includegraphics[scale=0.280,trim=10 0 100 60,clip]{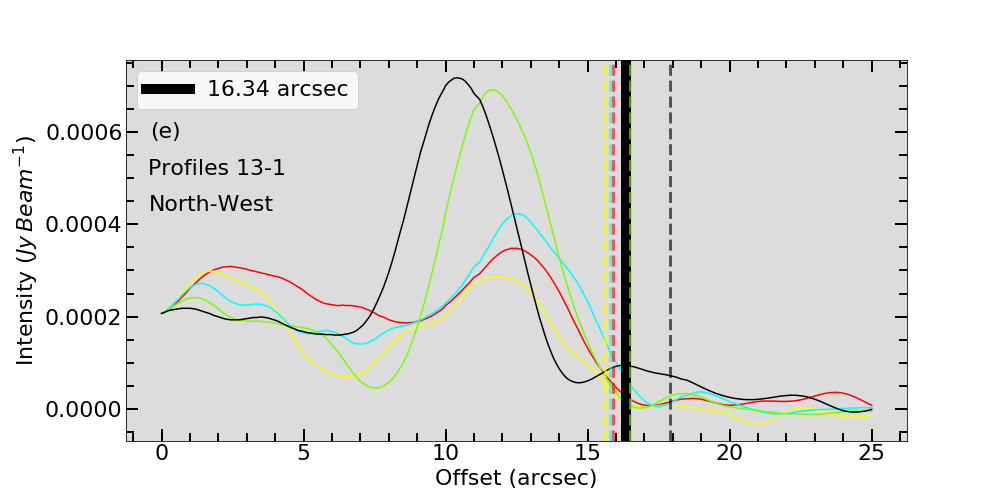}\\
\caption{Estimate of the radio continuum radius of \snr\ at 5500\,MHz. a: Radial profiles around the remnant from the centre (see Section~\ref{radio_mor}) overlaid on the $3\sigma$ \ac{ATCA} image contour (60\,$\mu$Jy\,beam$^{-1}$) at 5500\,MHz. The central position of the remnant is RA~(J2000)~=~05$^{h}$19$^{m}$34.85$^{s}$, Dec~(J2000)~=~$-$69$^\circ$02$'$08.22$''$. b, c, d, and e: Radial profiles. The dashed vertical lines represent profile cutoffs and the thick black vertical lines represent the average of the cutoffs (in arcsec) for different parts of the shell; south-west, south-east, and north-east, and north-west, respectively. 
\label{fig:slices}}
\end{figure*}

The resulting calculated centre is RA~(J2000)~=~05$^{h}$19$^{m}$34.85$^{s}$, Dec~(J2000)~=~$-$69$^\circ$02$'$08.22$''$ with an uncertainty of $\sim$0.1\,arcsec. This is $\sim$0.32\,$''$ ($\sim$0.07\,pc at the distance of 50\,kpc) south-west from the previous estimation by \cite{2012SerAJ.185...25B} (RA(J2000)~=~05$^{h}$19$^{m}$34.9$^{s}$, Dec~(J2000)~=~--69$^\circ$02$'$07.9$''$). 
To calculate the radio continuum radius of this \ac{SNR}, we used the \textsc{miriad} task \textsc{cgslice}  to plot 16 equispaced radial profiles in 22.5$^\circ$ segments around the remnant at 5500\,MHz. Each profile is 25$''$ in length (Figure~\ref{fig:slices}). We divide the remnant into four equal regions: south-west (profiles 1--5), south-east (profiles 5--9), north-east (profiles 9--13), and north-west (profiles 13--1) (Figure~\ref{fig:slices}a). We identify the cutoff as the point where each profile intersects the outer contour line ($3\sigma$ \ac{ATCA} image contour or 60\,$\mu$Jy\,beam$^{-1}$ at 5500\,MHz). These cutoffs are represented as dashed vertical lines (see Figure~\ref{fig:slices}b, c, d, and e). The thick black vertical lines represent the average of the cutoffs in each region. The resulting radii vary from $16.46\pm1.19''$ ($3.99\pm0.28$\,pc) towards the south-west (Figure~\ref{fig:slices}b) to $16.08\pm1.14''$ towards the south-east ($3.89\pm0.27$\,pc; Figure~\ref{fig:slices}c) and $18.44\pm0.75''$ ($4.47\pm0.18$\,pc) towards the north-east (Figure~\ref{fig:slices}d) to $16.34\pm0.93''$ ($3.96\pm0.22$\,pc) towards the north-west (Figure~\ref{fig:slices}e) with an average of $16.83\pm1.08''$ ($4.08\pm0.26$\,pc) for the entire \ac{SNR}. This is consistent with a previous estimation by \citet{2012SerAJ.185...25B}. The size of this \ac{SNR} is close to \acp{SNR} of similar age, J0509–673 \citep{2014MNRAS.440.3220B} in the \ac{LMC} and Kepler \citep{2012ApJ...756....6P} in the \ac{MW}.

%%%%%%%%%%%%%%%%%%%%%%%%%%%%%%%%%%%%%%%%%%%%%%%%%%%%%%%%%%%%%%%%%%%%%%%%%%%%%%

%

\subsection{Polarisation}
\label{pol}

The fractional polarisation for \snr\ was calculated using the Equation:

\begin{equation}
\label{eq1}
P=\frac{\sqrt{S^2_Q+S^2_U}}{S_I} ~,
\end{equation}
where $P$ is the average fractional polarisation, and $S_{Q}$, $S_{U}$, and $S_{I}$ are integrated intensities for the $Q$, $U$, and $I$ stokes parameters, respectively. 

\begin{figure*}[hbt!]
\centering
%\hspace{-0.42\textwidth}a)\hspace{0.47\textwidth}b) \\
\includegraphics[width=0.34\textwidth,angle=-90,trim=0 13 48 0,clip]{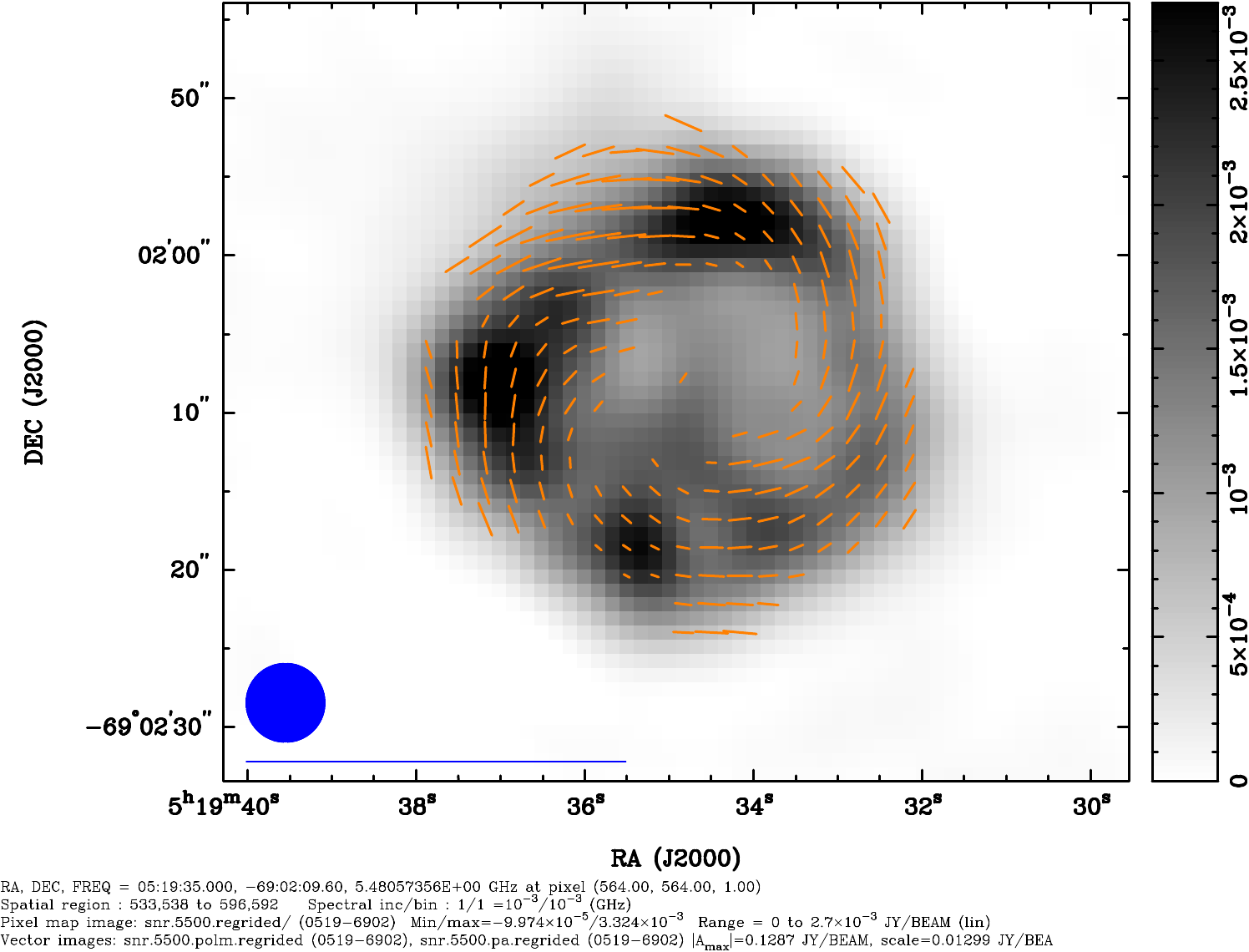}
\includegraphics[width=0.34\textwidth,angle=-90,trim=0 13 48 0,clip]{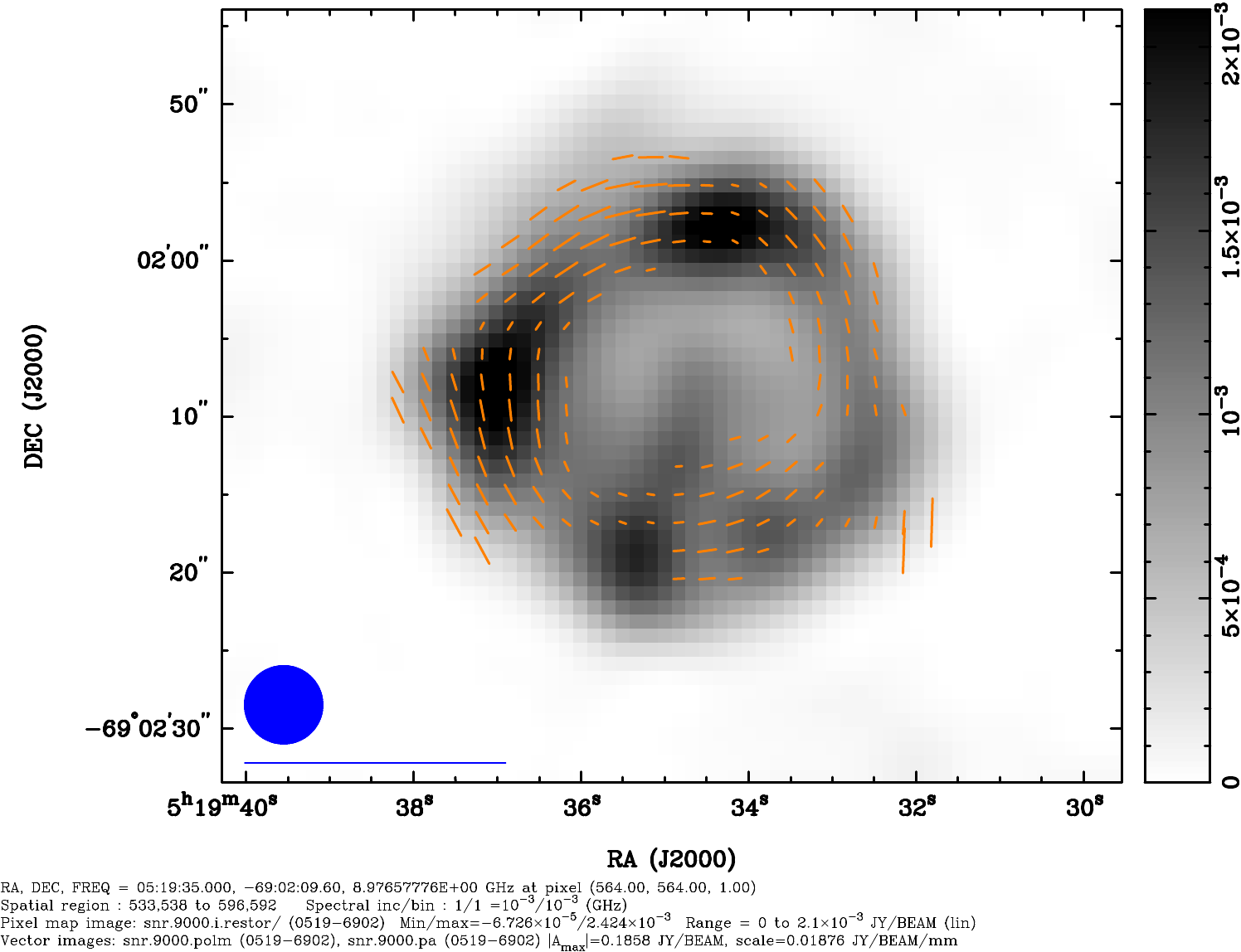}\\
%\hspace{-0.42\textwidth}c)\hspace{0.47\textwidth}d) \\
\includegraphics[width=0.34\textwidth,angle=-90,trim=0 0 0 0,clip]{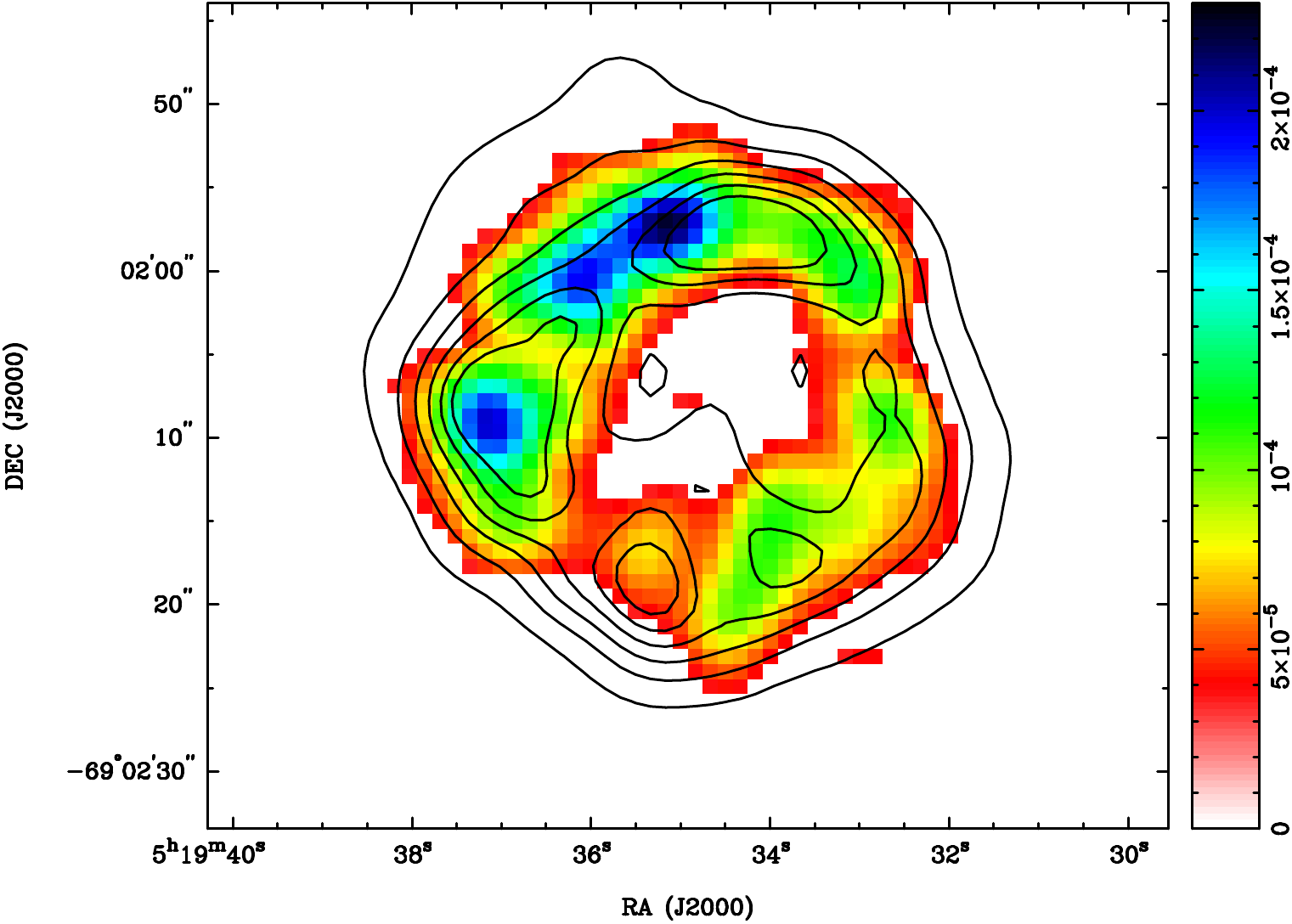}
\includegraphics[width=0.34\textwidth,angle=-90,trim=0 0 0 0,clip]{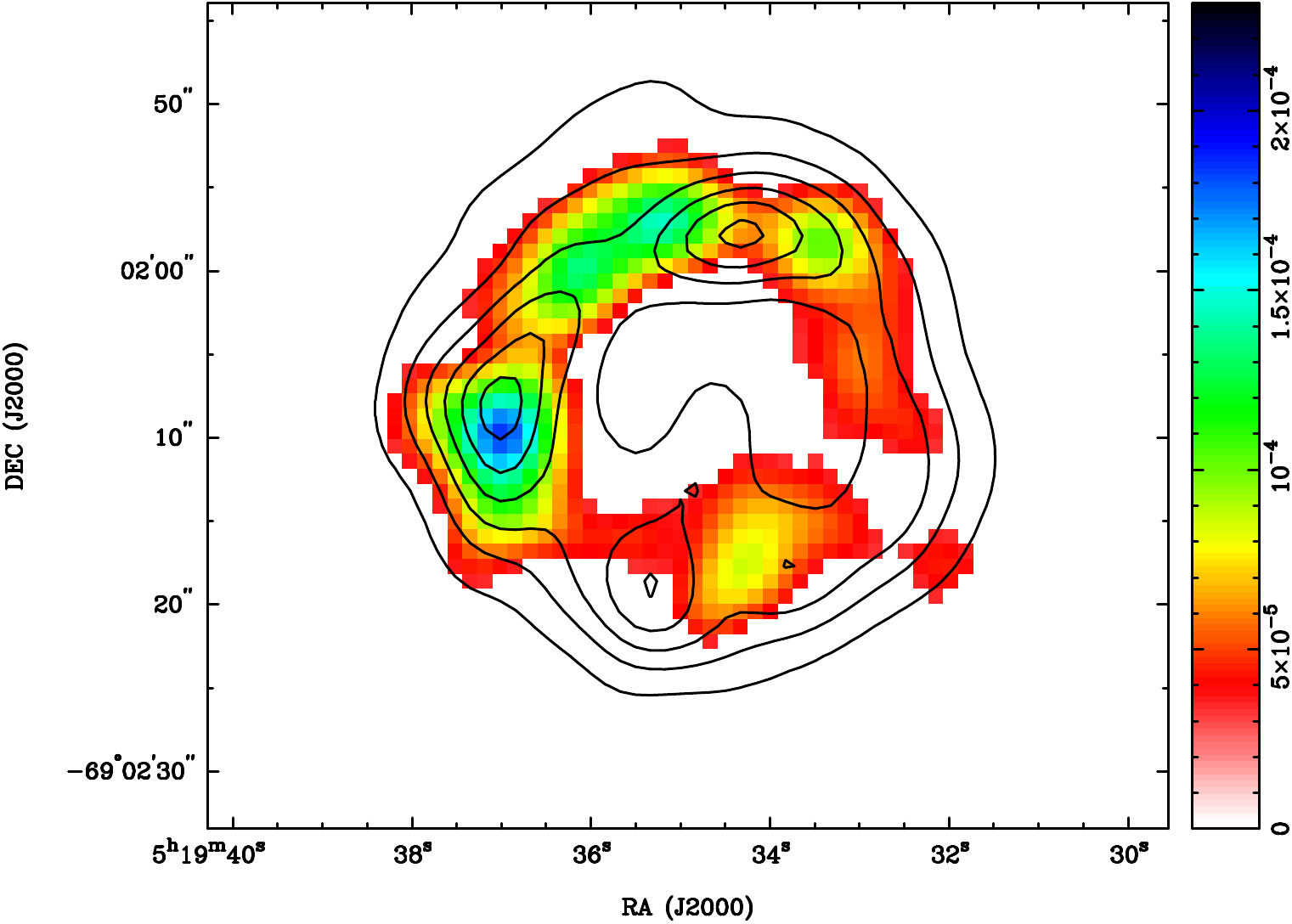}\\
\caption{Fractional Polarisation vectors of \snr\ overlaid on the intensity \ac{ATCA} images at 5500\,MHz (upper left) and at 9000\,MHz (upper right). The blue circle in the lower left corner represent a synthesised beam of $5''\times5''$ and the blue line below the circle represents 100\% polarisation. The bar on the right side represents the greyscale of the intensity images in Jy\,beam$^{-1}$. Polarisation intensity maps of \snr\ at 5500\,MHz (bottom left) and at 9000\,MHz (bottom right) are shown at bottom with intensity image contour lines overlaid. Respective contour levels are 0.2, 0.6, 1, 1.4, 1.8, and 2.2\,mJy\,beam$^{-1}$ for both images. The colour bar represents gradients of polarisation intensity.}
\label{fig:pol.5500}
\end{figure*}

\snr\ fractional polarisation vectors appear prominent in 5500 and 9000\,MHz maps (see Figure~\ref{fig:pol.5500}), we also present polarisation intensity maps in the same figure. The average fractional polarisation values are $5\pm1$\% and $6\pm1$\% for 5500 and 9000\,MHz respectively. These are fractionally lower than the values reported by \cite{2012SerAJ.185...25B} but consistent within the errors, and higher than the values of low frequencies (1472 and 2368\,MHz) reported by \cite{1995AJ....109..200D}. They are similar to the values of N\,103B of $\sim$8\% at 5500\,MHz \citep{2019Ap&SS.364..204A} in the \ac{LMC}, Kepler of $\sim6$\% at 4835\,MHz \citep{2002ApJ...580..914D}, and G1.9+0.3 of 6\% at 5500\,MHz \citep{2014SerAJ.189...41D} in the \ac{MW}.

%{\bf Why don't you talk here about the directions of polarisation vectors and connect that with typical for young SNRs?? }

%towards the north-east, north, and north-west parts of the remnant. The west and south-west parts are somewhat appear tangential at 9000\,MHz. While, there is a slightly polarisation at 5500\,MHz towards south and east regions. Figure.~\ref{fig:pol.5500} shows fractional polarisation maps at 5500 and 9000\,MHz, as well as polarisation intensity maps at each frequencies. We estimate a mean polarisation across the remnant of 10$\pm$2.7\% and 11$\pm$2.2\% for 5500 and 9000\,MHz images, respectively, which are higher than the values of a previous study by \cite{2012SerAJ.185...25B}. 

%The average values of polarisation intensity towards these regions are $\sim$40\expo{-6} and $\sim$27\expo{-6}, for 5500 and 9000\,MHz images, respectively.

%%%%%%%%%%%%%%%%%%%%%%%%%%%%%%%%%%%%% Fig 4 %%%%%%%%%%%%%%%%%%%%%%%%%%%%%%%%%%%%%%%%%%%

%%%%%%%%%%%%%%%%%%%%%%%%%%%%%%%%%%%%%%%%%%%%%%%%%%%%%%%%%%%%%%%%%%%%%%%%%%%%%%%%%%%%%
%

\subsection{Spectral Index}
\label{SI}

The radio spectra of \snr\ can be described as a pure power-law of frequency.

%$S_{\nu}$~$\propto$~$\nu^\alpha$, where $S_{\nu}$ is  flux density, $\nu$ is  frequency, and $\alpha$ is the spectral index. 

We used the \textsc{miriad} \citep{1995ASPC...77..433S} task \textsc{imfit} to extract a total integrated flux density from all available radio continuum observations of \snr\ listed in Table~\ref{tab2}. 
This includes observations from the \ac{MWA} \citep{2018MNRAS.480.2743F}, \ac{MOST} \citep{1976AuJPA..40....1C,2003MNRAS.342.1117M}, \ac{ASKAP} \citep{2021MNRAS.506.3540P}, and \ac{ATCA}. We measured the \ac{MWA} flux density for each sub-band (76--227\,MHz) and we also re-measured the \ac{ASKAP} flux density at 888\,MHz from \citet{2021MNRAS.506.3540P}. All arrays are sensitive to angular scales much larger than the size of the \ac{SNR}. Therefore, the sampling of the $uv$ plane for different arrays and at different frequencies would not affect flux density measurements of extended structures. 
For cross-checking and consistency, we used \textsc{aegean} \citep{hancock2018} and found no significant difference in integrated flux density estimates. 
Namely, we measured \snr\ local background noise (1$\sigma$) and carefully selected the exact area of the \ac{SNR} which also excludes all obvious unrelated point sources. We then estimated the sum of all brightnesses above 5$\sigma$ of each individual pixel within that area and converted it to \ac{SNR} integrated flux density following \citet[][Equation~24]{1966ARA&A...4...77F}. 
We also estimate that the corresponding radio flux density errors are below 10\,per\,cent as examined in our previous work \citep{2022MNRAS.512..265F,2023MNRAS.518.2574B}. %In this estimate, various contributions to the flux density error are considered, including missing short spacings. 

%Although for weaker sources this uncertainty is more proclaimed, for brighter objects such as our SNR E0102, the flux density error is much smaller ($<$ 10 per,cent).}

\begin{figure}
 \begin{center}
% %\includegraphics[width=\columnwidth]{fig1.eps}
\includegraphics[scale=0.45,trim=10 10 20 40,clip]{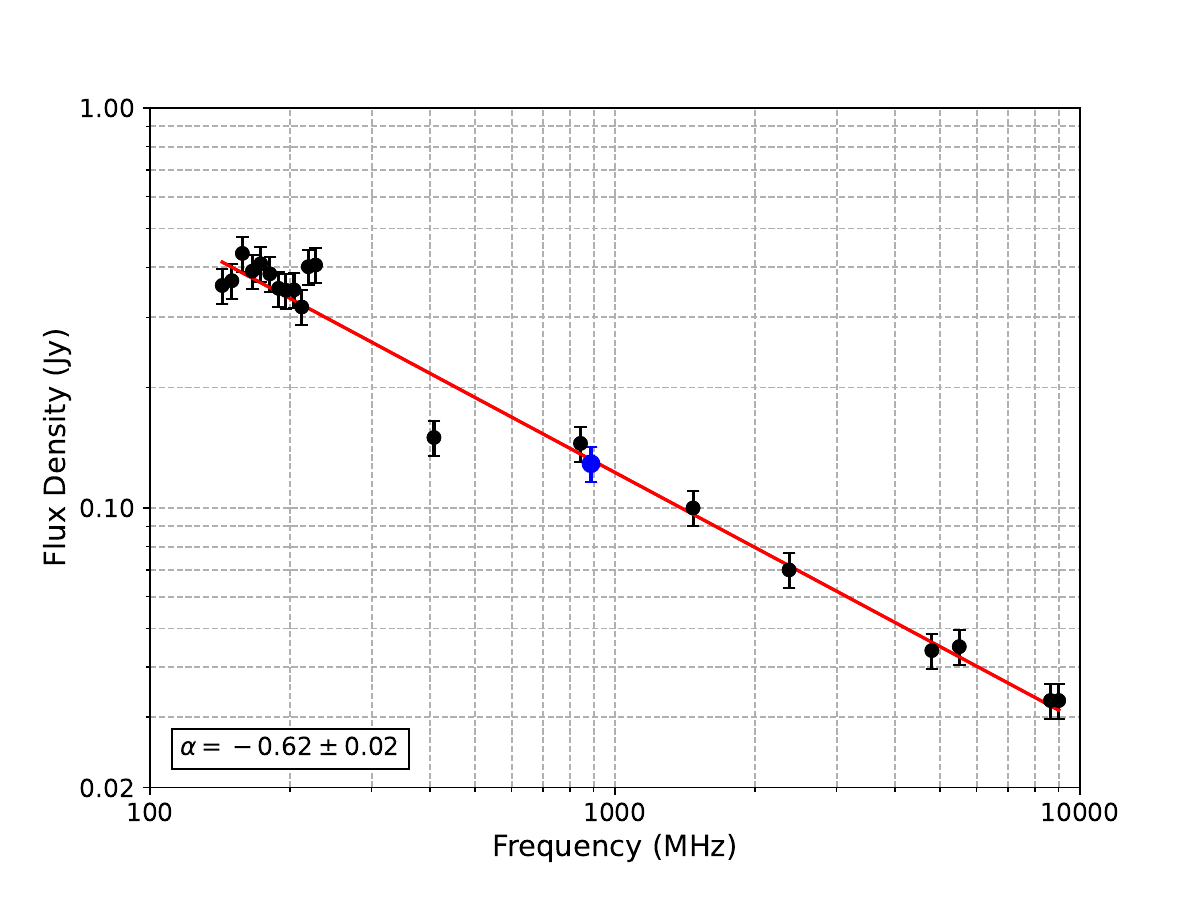}\\
 \caption{Radio continuum spectrum of \snr. The blue dot indicates that we re-measured this flux density.}\label{Fig8}
\end{center}
 \end{figure}

In Figure~\ref{Fig8}, we present the flux density vs. frequency graph for \snr.
The relative errors are used for the error bars on a logarithmic plot.
The best power-law weighted least-squares fit is shown (thick red line), with the spatially integrated spectral index $\langle\alpha\rangle= {-0.62 \pm 0.02}$; slightly steeper but within  range when compared to \cite{2012SerAJ.185...25B}. This spectral index is consistent with values of similar aged \acp{SNR} such as Kepler \citep[--0.64,][]{1988ApJ...330..254D} and SN\,1006 \citep[--0.6,][]{1965AJ.....70..754G} within the \ac{MW}.  
%\Denis{ not sure what you mean by the above: is it general for all SNRs or has the multifrequency spectrum of this SNR been measured before with no indication of spectral curvature?}

\begin{figure*}
\begin{center}
\includegraphics[scale=0.35,angle=-90,trim=0 0 0 0,clip]{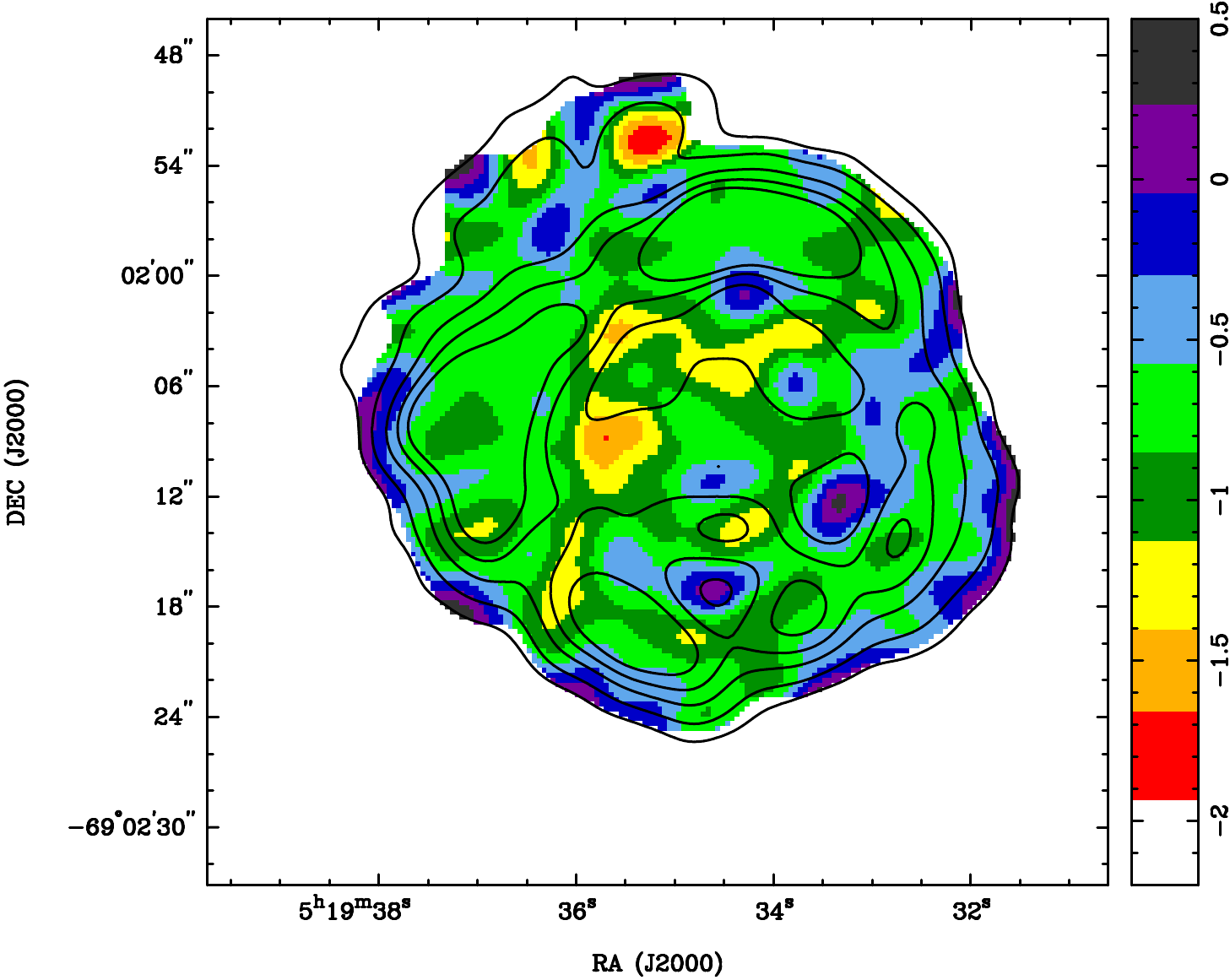}
\includegraphics[scale=0.35,angle=-90,trim=0 0 0 0,clip]{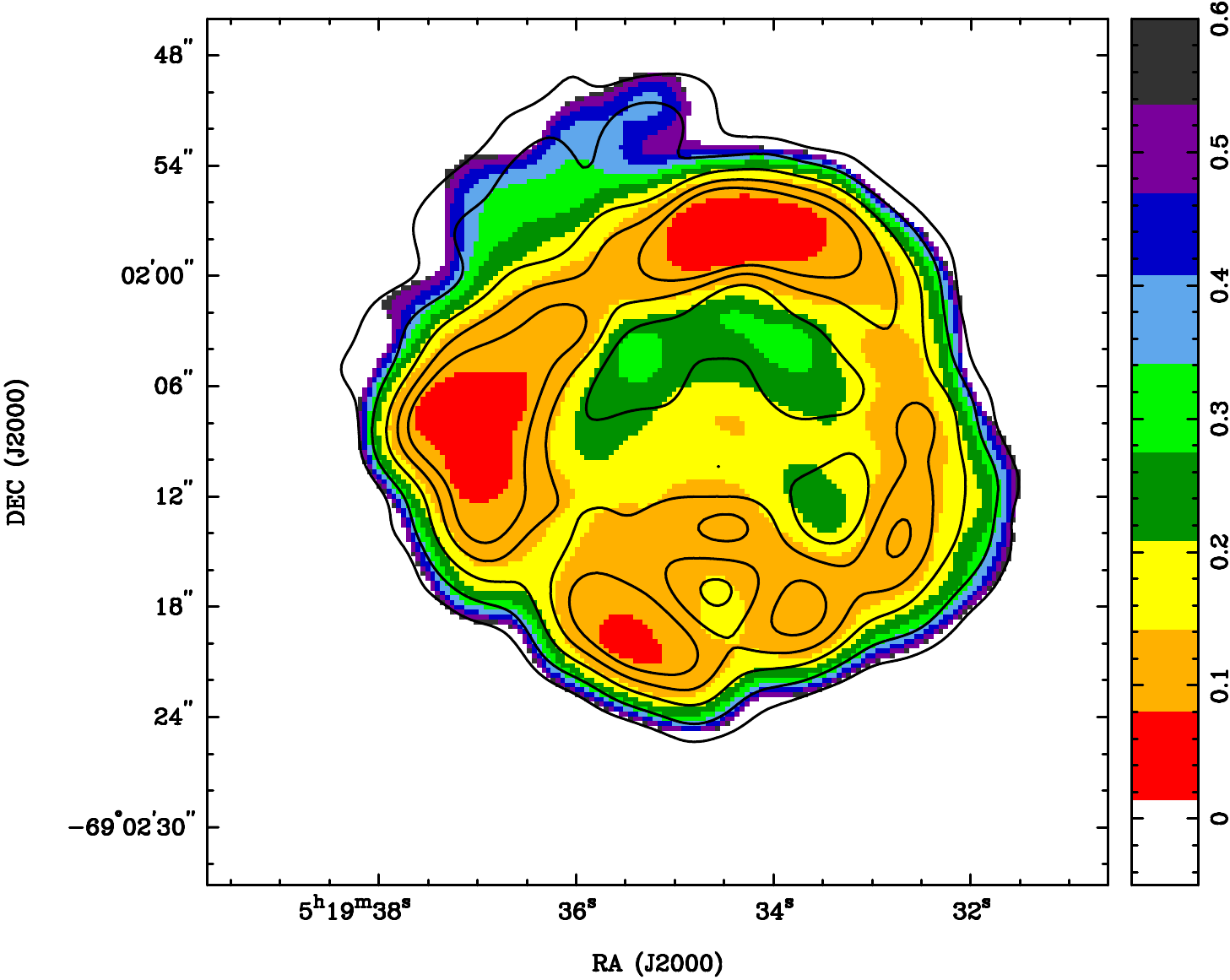}\\
\caption{\emph{Left}: Spectral index map of \ac{SNR} \snr\ created using 5500 and 9000\,MHz images. \emph{Right}: Error map of spectral index of \ac{SNR} \snr. Contour lines are overlaid using our \ac{ATCA} 5500\,MHz image (0.09, 0.2, 0.4, 0.6, and 0.8\,mJy\,beam$^{-1}$).}
\label{Fig7}
\end{center}
\end{figure*}

We also produced a spectral index map using 5500 and 9000\,MHz images. To create this, all images were first re-gridded to the finest pixel size ($0.26''\times0.26''$) using the \textsc{miriad} task \textsc{regrid}. The images were then smoothed to a common resolution ($3''\times3''$) using \textsc{convol}. Finally, \textsc{maths} was applied to create the spectral index map and its corresponding error map as shown in Figure~\ref{Fig7}. The average spectral index value across the remnant was found to be ${-0.7\pm0.2}$. We note that the uncertainty value is an order of magnitude higher than the estimate from Figure~\ref{Fig8}. This is because the error map of spectral index was generated using only two images (5500 and 9000\,MHz).
%DL: I don't see the following: the bright region in N has -0.6, thus better to omit.
%We note that the spectral index is steep ($\alpha\sim-0.8$) where the radio continuum emissions peak, and becomes even steeper ($\alpha\sim-1$) for the surrounding regions -- with the steepest ($\alpha\sim-1.3$) region towards the centre of the remnant (Figure~\ref{Fig7}). 
%DL: suggest to omit the following: you can't calculate a reliable spectral index for the faintest regions. Did you make a spectral index error map?
%The inner and outer radii show a flat spectral index ($\alpha\sim-0.2$) with a flatter ($\alpha\sim0$) region towards the north side of the remnant where there is no radio continuum emissions (see Figure~\ref{Fig7}).

%{\bf Why don't you compare here this snr with others of similar age and type (type~Ia)?? }

%%%%%%%%%%%%%%%%%%%%%%%%%%%%%%% Fig 5 %%%%%%%%%%%%%%%%%%%%%%%%%%%%%%%%%%%%%%%%%%%%%

%%%%%%%%%%%%%%%%%%%%%%%%%%%%%%%%%%%%%%%%%%%%%%%%%%%%%%%%%%%%%%%%%%%%%%%%%

%%%%%%%%%%%%%%%%%%%%%%%%%%%%%%%%%% Fig 6 %%%%%%%%%%%%%%%%%%%%%%%%%%%%%%%%%%%%
\begin{table}[ht!]
\caption{Flux density measurements of \snr. The asterisk ($^{*}$) indicates that we re-measured this flux density.}
\footnotesize
\centering
\begin{tabular}{@{}ccccc@{}}
\hline\hline
Freq.     &    Flux Density &  Telescope & Reference\\
(MHz)     &     (Jy)  &         \\
 \hline%
143           &         0.360  &       MWA        &  This work\\
150           &         0.370  &       MWA        &  This work\\
158           &         0.433  &       MWA        &  This work\\
166           &         0.391  &       MWA        &  This work\\
173           &         0.408  &       MWA        &  This work\\
181           &         0.385  &       MWA        &  This work\\
189           &         0.354  &       MWA        &  This work\\
196           &         0.350  &       MWA        &  This work\\
204           &         0.351  &       MWA        &  This work\\
212           &         0.318  &       MWA        &  This work\\
219           &         0.401  &       MWA        &  This work\\
227           &         0.405  &       MWA        &  This work\\
408           &0.150  &       MOST       & \cite{2012SerAJ.185...25B} \\
843           & 0.145 &       MOST       & \cite{2012SerAJ.185...25B}\\
888           &   0.129$^{*}$  &      ASKAP       & \cite{2021MNRAS.506.3540P} \\
1472          &         0.100  &       ATCA       & \cite{1995AJ....109..200D}\\
2368 & 0.070 &       ATCA       & 
%be very careful here, you had 2378 but I think you meant 2378, double check}
\cite{1995AJ....109..200D}\\
%2400      &    0.064  &    0.0064      &ATCA       & \cite{2012SerAJ.185...25B} \\
4800          &    0.044      &    ATCA           & \cite{2017ApJS..230....2B}\\
5500          &    0.045      &    ATCA           & This work\\
8640          &    0.033      &    ATCA           & \cite{2017ApJS..230....2B}\\
9000          &    0.033      &    ATCA           & This work\\
\hline\hline
\end{tabular}
\label{tab2}
\end{table}

\subsection{Rotation Measure and Magnetic Field}
\begin{figure}[ht!]
\begin{center}
\includegraphics[scale=0.32,angle=270,trim=0 0 5 0,clip]{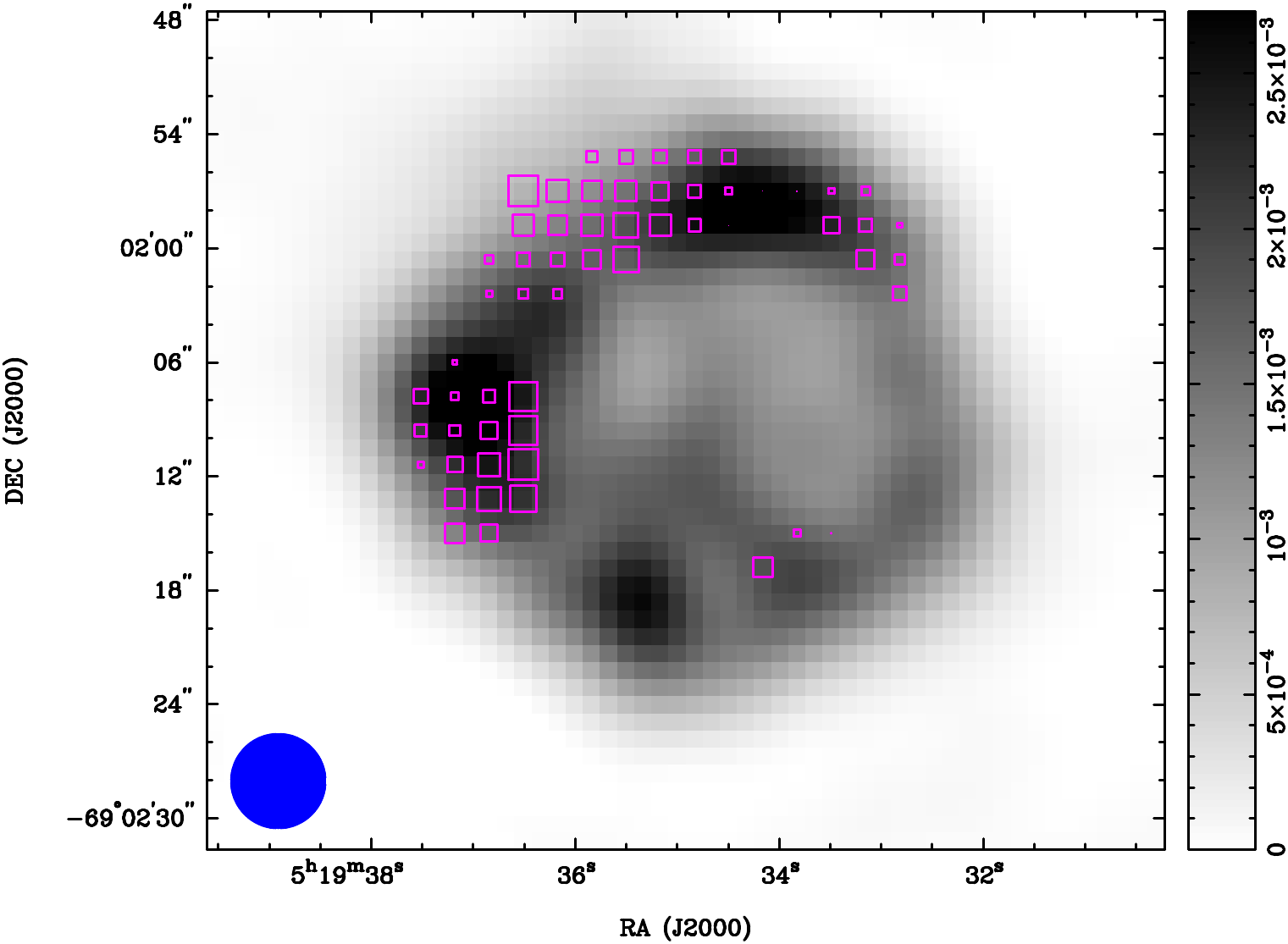}\\
%\hspace{-0.42\textwidth}a)\hspace{0.55\textwidth}b) \\
%\includegraphics[width=0.375\textwidth,angle=270,trim=0 0 0 0,clip]{figures/RM_3-3.eps}
%\includegraphics[width=0.37\textwidth,angle=270,trim=0 0 0 0,clip]{figures/RM_3-3_N-W.eps}\\
%\hspace{-0.42\textwidth}c)\hspace{0.55\textwidth}d) \\
%\includegraphics[width=0.335\textwidth,angle=270,trim=0 0 0 0,clip]{figures/RM_3-3_N.eps}
%\includegraphics[width=0.335\textwidth,angle=270,trim=0 0 0 0,clip]{figures/RM_3-3_S-E.eps}\\
\caption{Rotation measure boxes of \snr\ (\ac{ATCA}; 4723, 5244, 5756, and 6268\,MHz) overlaid on the \ac{ATCA} image at 5500\,MHz (greyscale). Most boxes are open (negative values), ranging from a minimum of --314\,rad\,m$^{-2}$ to a maximum of 33\,rad\,m$^{-2}$. The blue circle in the lower left corner represents a synthesised beam of $5''\times5''$. The bar on the right side represents the greyscale of the 5500\,MHz \ac{ATCA} image in Jy\,beam$^{-1}$.}
\label{RM}
\end{center}
\end{figure}

\begin{figure}[ht!]
\begin{center}
\includegraphics[scale=0.32,angle=270,trim=40 10 23 40,clip]{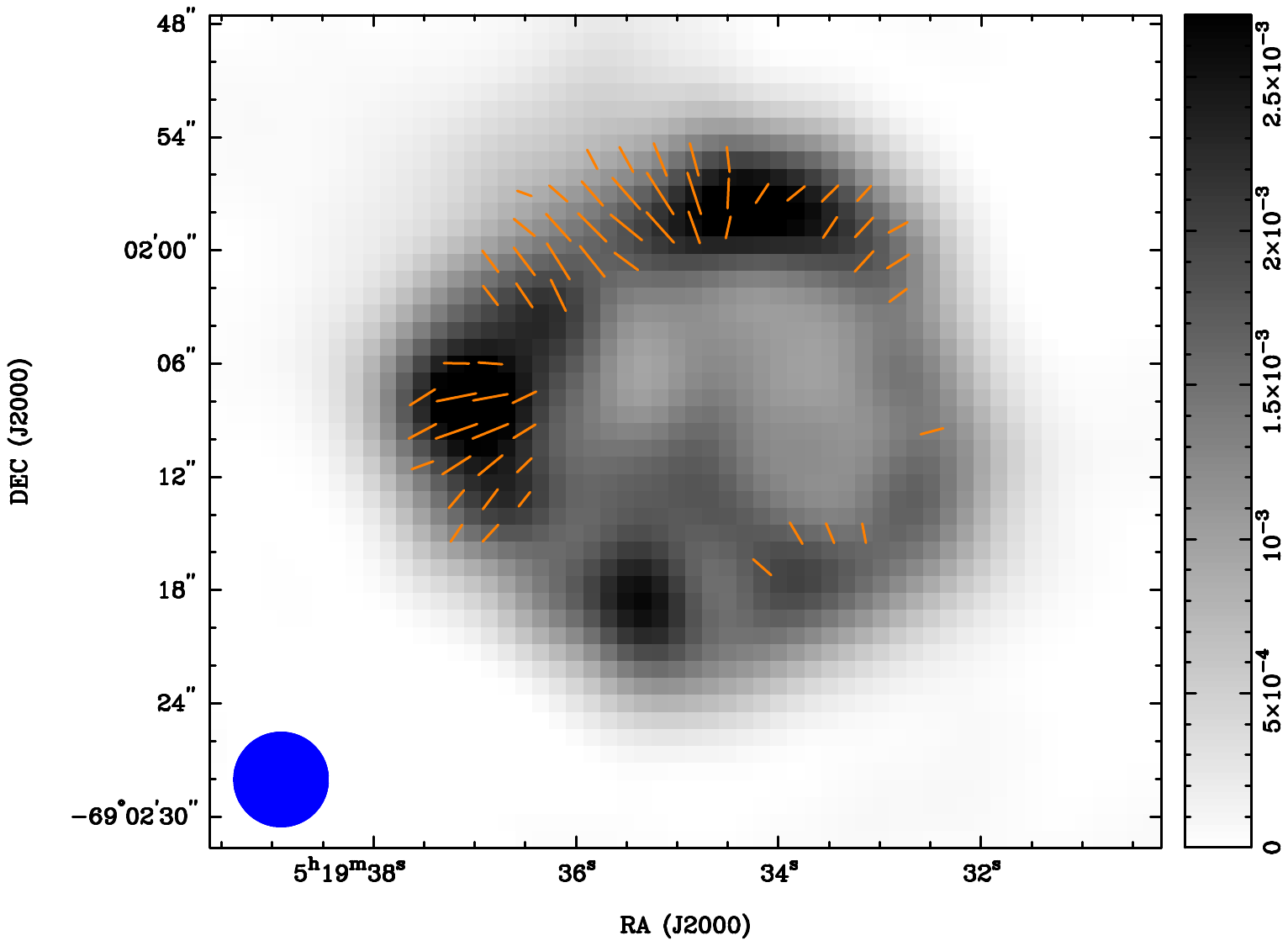}\\
%\hspace{-0.42\textwidth}a)\hspace{0.55\textwidth}b) \\
%\includegraphics[width=0.375\textwidth,angle=270,trim=0 0 0 0,clip]{figures/RM_3-3.eps}
%\includegraphics[width=0.37\textwidth,angle=270,trim=0 0 0 0,clip]{figures/RM_3-3_N-W.eps}\\
%\hspace{-0.42\textwidth}c)\hspace{0.55\textwidth}d) \\
%\includegraphics[width=0.335\textwidth,angle=270,trim=0 0 0 0,clip]{figures/RM_3-3_N.eps}
%\includegraphics[width=0.335\textwidth,angle=270,trim=0 0 0 0,clip]{figures/RM_3-3_S-E.eps}\\
\caption{Magnetic field direction vectors of \snr\ overlaid on the \ac{ATCA} image at 5500\,MHz (greyscale). The blue circle in the lower left corner represents a synthesised beam of $5''\times5''$. The bar on the right side represents the greyscale of the 5500\,MHz \ac{ATCA} image in Jy\,beam$^{-1}$.}
\label{MFD}
\end{center}
\end{figure}

To calculate the \ac{RM} of \snr, the 2048\,MHz bandwidth at 5500\,MHz was split into four 512\,MHz sub-bands (4723, 5244, 5756, and 6268\,MHz). Using position angle measurements associated with fractional polarisation maps for these frequencies, the \textsc{miriad} task \textsc{imrm} was applied. \ac{RM} values with an error $\geq$~120\,rad\,m$^{-2}$ were flagged. The resulting \ac{RM} map is shown in Figure~\ref{RM}. We note  the \ac{RM} values are mostly negative with some positive values towards the north and west of the remnant. Values of \ac{RM} are concentrated in three areas of the remnant; north-east, north, and north-west (see Figure~\ref{RM}), with an average value of ${-124\pm83}$\,rad\,m$^{-2}$, different than the value reported by \cite{2012SerAJ.185...25B}. Once the \ac{RM} had been determined, the vectors were rotated back to their intrinsic (zero wavelength) values. In Figure~\ref{MFD} an additional 90$^\circ$ was added to the electric vectors so that the vectors represent the direction of magnetic-fields within the remnant. These vectors are oriented in a radial direction (see Figure~\ref{MFD}), as expected from a young \ac{SNR}.

To calculate the \ac{ED} for \snr, we used Equation~\ref{eq2}, where $EM$ is the emission measure in cm$^{-3}$, $D$ is the distance to the \ac{SNR}; \ac{ED} and $n_p$ are electron and proton densities, respectively.

\begin{equation}
 \label{eq2}  
 EM = \int \ac{ED}*n_p~dV/(4\pi D^2) ~,
\end{equation}

\noindent By assuming \ac{ED} = 1.2\,$n_p$, a volume filling factor = 1, an isotropic and spherical distribution of ionised gas, and using an $EM\sim 22.5 \times 10^{58}$\,cm$^{-3}$
%DL from Maggi+2016 Table E2 the EM is 22.5x10^{58}: did you use the wrong value?
%Rami I used the value in table E1 which is 4.53
\citep{2016A&A...585A.162M}, we obtain \ac{ED} = 6.8\,cm$^{-3}$.

The magnetic field strength can be estimated using the Equation for Faraday depth\footnote{This assumes only one source along the line of sight with no internal Faraday rotation; the Faraday depth approximates the rotation measure of all wavelengths.}:

\begin{equation}
    \label{eq3}
    RM=811.9\int_{0}^{L}n_{e}B_{\parallel}dl ,
\end{equation}

\noindent where \ac{RM} is the rotation measure in rad\,m$^{-2}$, \ac{ED} is the electron density in cm$^{-3}$, $B_{\parallel}$ is the line of sight magnetic field strength in $\mu$G, and $L$ is the path length through the Faraday rotating medium in kpc \citep{2004JKAS...37..337C}. 

Using our average value of \ac{RM} (--124\,rad\,m$^{-2}$), an \ac{SNR} thickness ($L$) of the compressed shell of $\sim$2\,pc, and an \ac{ED} value of 6.8\,cm$^{-3}$, we obtain an average magnetic field strength of 11.2\,$\mu$G.

The equipartition model\footnote{Online calculator is found at \url{http://poincare.matf.bg.ac.rs/~arbo/eqp}} can also be used to estimate the magnetic field strength for \snr. This method uses modeling and simple parameters to estimate the intrinsic magnetic field strength and energy contained in the magnetic field and cosmic ray (CR) particles using radio synchrotron emission \citep{2012ApJ...746...79A,2013ApJ...777...31A,2018ApJ...855...59U}.

This approach is purely analytical, described as a rough, only order of magnitude estimate because of assumptions used in analytical derivations and errors in the determination of distance, angular diameter, spectral index, a filling factor, and flux density, tailored especially for the magnetic field strength in SNRs. \citet{2012ApJ...746...79A,2013ApJ...777...31A,2018ApJ...855...59U} present two models; the difference is in the assumption whether there is equipartition, precisely constant partition with CRs or only CR electrons. \citet{2018ApJ...855...59U} showed the latter type of equipartition is a better assumption than the former.

Using the \citet{2018ApJ...855...59U} model\footnote{We use: $\alpha=0.62$, $\theta$~=~0.26\,arcmin, $\kappa=0$, S$_{\rm 5.5\,GHz}$~=~0.045\,Jy, and f~=~0.87.}, the mean equipartition field over the entire remnant is ${72\pm5}$\,$\mu$G with an estimated minimum energy of E$_{\rm min}$~=~2.6$\times10^{48}$\,erg. However, using the \citet{2012ApJ...746...79A} earlier model, which assumes CRs composed of electrons, protons, and ions, the estimate is ${156\pm5}$\,$\mu$G, with a minimum explosion energy of E$_{\rm min}$~=~1.2$\times10^{49}$\,erg. This is consistent with previous values reported by \cite{2012SerAJ.185...25B}.

Of course, we can not suppose we know the remnant's magnetic field to the significant figures implied above as the models do not have any such accuracy. We do feel it is reasonable to say \snr's magnetic field can be estimated between 10 and $10^2$\,$\mu$G. This is consistent with other relatively young Type\,Ia \acp{SNR} including Tycho, Kepler, and SN\,1006 \citep{1992ApJ...399..L75}.

%Let's stay out of the kappa thing, there no reason to go into it, instructions are on the website ;-)
%%%%%%%%%%%%%%%%%%%%%%%%%%%%%%%%%%%%% Fig. 7 %%%%%%%%%%%%%%%%%%%%%%%%%%%%%%%%%%%%%%%%%%%%
%\includegraphics[width=\columnwidth]{fig1.eps}
%%%%%%%%%%%%%%%%%%%%%%%%%%%%%%%%%%%%%%%%%%%%%%%%%%%%%%%%%%%%%%%%%%%%%%%%%%%%%%%%%%%

%%%%%%%%%%%%%%%%%%%%%%%%%%%%%%%%%%%%% Fig. 8 %%%%%%%%%%%%%%%%%%%%%%%%%%%%%%%%%%%%%%%%%%%%
%\includegraphics[width=\columnwidth]{fig1.eps}

\subsection{$\Sigma$--D Relation}
\begin{figure}[ht!]
\begin{center}
\includegraphics[width=\columnwidth,trim=0 0.9 0 0,scale=1.5,clip]{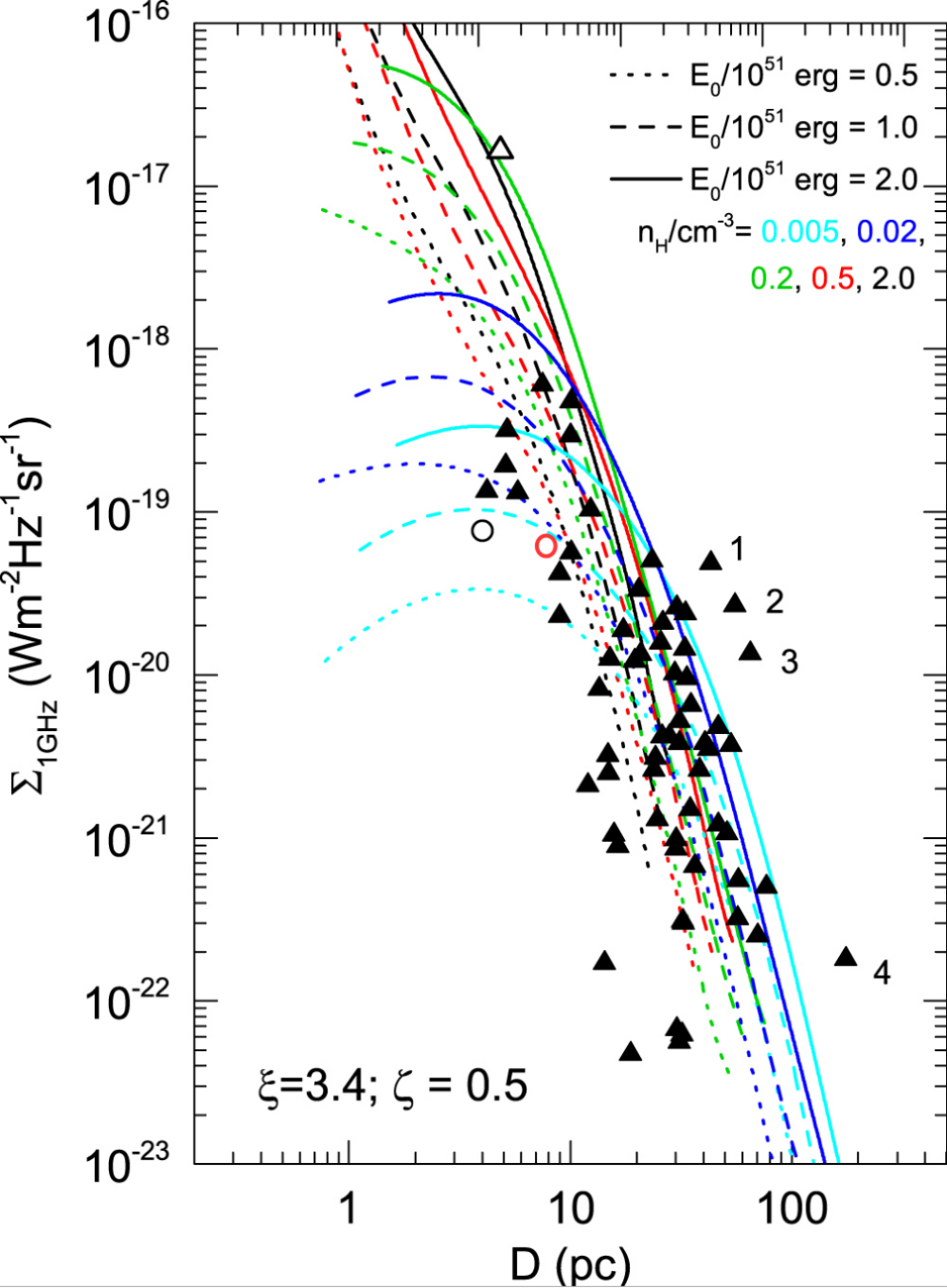}\\
%\hspace{-0.42\textwidth}a)\hspace{0.55\textwidth}b) \\
%\includegraphics[width=0.375\textwidth,angle=270,trim=0 0 0 0,clip]{figures/RM_3-3.eps}
%\includegraphics[width=0.37\textwidth,angle=270,trim=0 0 0 0,clip]{figures/RM_3-3_N-W.eps}\\
%\hspace{-0.42\textwidth}c)\hspace{0.55\textwidth}d) \\
%\includegraphics[width=0.335\textwidth,angle=270,trim=0 0 0 0,clip]{figures/RM_3-3_N.eps}
%\includegraphics[width=0.335\textwidth,angle=270,trim=0 0 0 0,clip]{figures/RM_3-3_S-E.eps}\\
\caption{Radio surface brightness–to–diameter diagram for \acp{SNR} at a frequency of 1\,GHz (black triangles), obtained from numerical simulations \citep[][their Figure~3]{2018ApJ...852...84P}. \snr\ is marked with an open red circle while the open triangle represents Cassiopeia\,A. The open black circle represents the youngest Galactic \ac{SNR}, G1.9+0.3 \citep{2020MNRAS.492.2606L}. Numbers represent the following \acp{SNR}: (1) CTB\,37A, (2) Kes\,97, (3) CTB\,37B, and (4) G65.1+0.6.}
\label{sigma_D}
\end{center}
\end{figure}

A previous $\Sigma$--D diagram \citep[][their Figure~3]{2022PASP..134f1001U,2020NatAs...4..910U,2018ApJ...852...84P} was compared with our values, D$\sim$8\,pc and $\Sigma_{\rm 1\,GHz} =6\times10^{-20}$\,W\,m$^{-2}$\,Hz$^{-1}$\,sr$^{-1}$, for \snr. This comparison implies the \snr\ is undergoing expansion within a surrounding environment characterised by a low-density of 0.005--0.02\,cm$^{-3}$ with an initial energy of explosion of E$_{\rm 0}$~=~$10^{51}$\,erg (Figure~\ref{sigma_D}). \snr\ position at $\Sigma$--D tracks can set it somewhere at the end of the free expansion phase, close to entering the early Sedov phase of evolution. 
%\newpage

Of course, multiple tracks can correspond to a single \ac{SNR} on the $\Sigma$--D plane. Assuming this is a young \ac{SNR}, it is unlikely that the black and red lines would apply. If the evolutionary tracks are black or red in this position on the $\Sigma$--D diagram, then this \ac{SNR} would need to be old, which does not fit. \cite{2018ApJ...852...84P} gives an explanation for 1-4 outliers. Alternatively, application of the method from \cite{2022PASP..134f1001U,2020NatAs...4..910U} to determine the evolutionary status of newly detected \acp{SNR}, the position on the $\Sigma$--D diagram (where a single \ac{SNR} might align with multiple tracks) must relate to the equipartition magnetic field strength and spectral characteristics. According to the literature, \snr\ is a young \ac{SNR}, and its position on the $\Sigma$--D diagram, equipartition strength, and steep spectrum (--0.62) support this classification. In this paper, the evolutionary phase using \cite{2022PASP..134f1001U,2020NatAs...4..910U} is not estimated; instead, assuming it is a young \ac{SNR}, environmental density is estimated based on the $\Sigma$--D tracks.

\subsection{Distribution of \HI\ Clouds} 
\label{HI}

\begin{figure}[ht!]
\begin{center}
\includegraphics[width=\columnwidth]{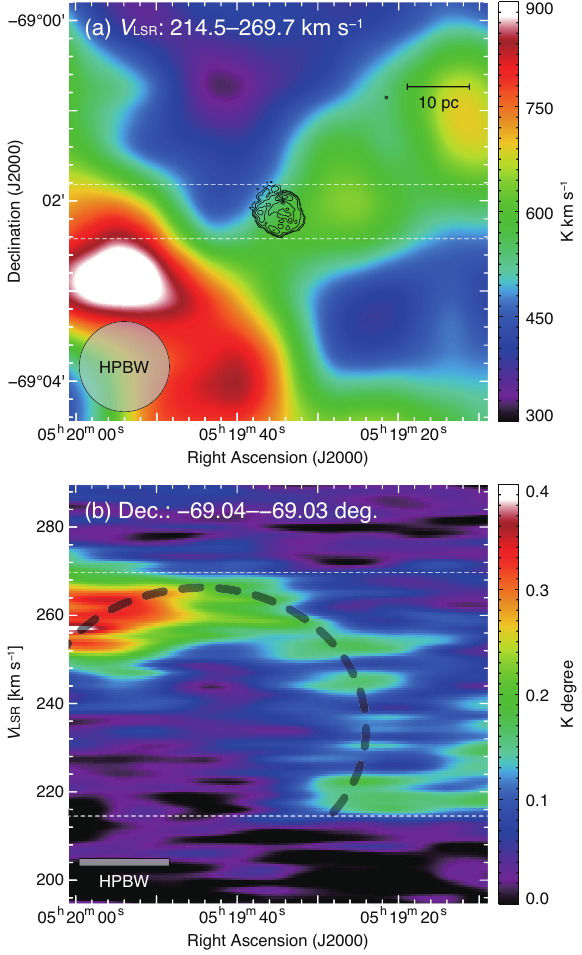}
\caption{(a) Integrated intensity map of \HI\ towards \snr. The integrated velocity range is from 214.5 to 269.7 km s$^{-1}$. The superposed contours are the same as shown in Figure \ref{Fig1}a. (b) Position--velocity diagram of \HI. The integrated Declination range is from $-$69.04$^{\circ}$ to $-$69.03$^{\circ}$. The dashed curve in the position--velocity diagram indicates the boundary of the \HI\ cavity (see Section~\ref{HI}).}
\label{Fig9}
\end{center}
\end{figure}

The distribution of \HI\ in the LMC velocity range %DL added
towards \snr\ is shown in Figure~\ref{Fig9}a where bright \HI\ clouds lie to the south-east. At this radial velocity from $\sim$220 to $\sim$270~km~s$^{-1}$ alone, \HI\ gas was predominantly present (see Appendix for details). However, we could not confirm an association because the angular resolution of the \HI\ data is roughly twice the diameter of the \ac{SNR}. The \ac{SNR} shell \HI\ integrated intensity is $\sim$600\,K\,km\,s$^{-1}$, corresponding to a column density of $\sim1 \times 10^{21}$\,cm$^{-2}$ under an optically thin assumption \citep[e.g.,][]{1990ARA&A..28..215D}.

Figure~\ref{Fig9}b shows a position--velocity diagram centred near the \ac{SNR} shell % 
%\Jeff{  Denis, just in process of revising and rewrote some of this, some might now make more sense assuming I understand what Rami is trying to say, but all your other points are well taken for Rami to address so I'm going to leave this section as is for now.} 
on the edge of an incomplete cavity-like structure of \HI\ that spans a diameter of $\sim$50\,pc and a velocity of $\sim$30\,km s$^{-1}$, respectively. \snr\ possibly exploded at the edge of what may be a wind bubble that might have originated from a fast outflow driven from an accreting \ac{WD}; similar to that proposed for Tycho \citep{2016ApJ...826...34Z,2021ApJ...906L...3T} and N\,103B \citep{2018ApJ...867....7S}. %However, Confirming this is challenging due to the low resolution of \HI\ data. 
%and favouring a \ac{SD} scenario for our \ac{SNR}. 
Using a model from \citet{2017ApJ...837...36L} having an updated radius of 3.6\,pc in a wind environment, we find new ages with the same model between 250 and 1100\,yrs for various values of the ejecta density profile index ($r^{-n}$ with $n=6$ to 14), compared to $\sim 2700$\,yrs for a uniform circumstellar medium.

%%%%%%%%%%%%%%%%%%%%%%%%%%%%%%%%%%%%% Fig. 9 %%%%%%%%%%%%%%%%%%%%%%%%%%%%%%%%%%%%%%%%%%%%
%%%%%%%%%%%%%%%%%%%%%%%%%%%%%%%%%%%%%%%%%%%%%%%%%%%%%%%%%%%%%%%%%%%%%%%%%%%%%%%%%%%%%%%%%%%

% \section{Comparison to Similar Age and Type\,Ia \acp{SNR}}
%  \label{sec:comparison}
 
% \hs{comments: I think MC~SNR~J0519$-$6902 is possibly exploded in the edge of the wind-bubble, but the origin of wind-bubble is still unknown (possibly an old supperbubble??). Also, I would appreciate that if you could refer some papers as references of the SD origin. Tycho's SNR: Zhou et al. 2016, ApJ, 826, 34; Tanaka et al. 2021, ApJL, 906, 3, N103B: Sano et al. 2018, ApJ, 867, 7.}

%{\color{red} Jeff: I don't really get the point of the following paragraph:}
At the same time, the tentative bubble may not originate from the progenitor star(s). Those winds eject of order one solar mass of material at low velocities with low energy.
%\Denis{ ok but need to compare below to estimate age and mass of HI shell} 
%It is certainly true that it would not be easy to can create
%DL I don't think this is justified.
The \HI\ bubble has a size of 30\,pc at the distance of the \ac{LMC}. We use \cite{2007pcim.book.....K} to calculate the radius and mass of the swept-up \HI\ shell. Using his Equation (16.62), we obtain the shell radius of 1.9\,pc for a wind speed of 20\,km\,s$^{-1}$, \ac{ISM} density of 1\,cm$^{-3}$ and wind duration of 1\,Myrs and 55\,pc for a duration of 100\,Myrs. The mass of the \HI\ shell would be 0.8\,M\textsubscript{\(\odot\)} and 2$\times 10^{4}$\,M\textsubscript{\(\odot\)} for duration of 1 and 100\,Myrs, respectively.
%\Denis{  need to discuss the size of bubble in terms of stellar wind parameters. e.g. Kwok Interstellar Medium textbook (University Science Books) section 16.3 equation 16.60}
The mass of the partial shell feature that is visible in \HI\ in Figure~\ref{Fig9}b is estimated by integrating the \HI\ over velocity and over the area of the shell (with radius $\sim$30\,pc). The resulting mass in the feature is $\sim$2.5 M$_{\odot}$, which is consistent with the theoretical estimate above if the duration of the wind is $\sim$3 million\,yrs. The radius of a shell expanding at 10\,km\,s$^{-1}$ for 3 million\,yrs is 30\,pc, consistent with a low-mass stellar wind origin of the shell. 
%\Denis{ need to calculate an estimated mass in the cavity wall}
In any case, further \HI\ studies with  higher angular resolutions
are needed to confirm this scenario for \snr.

\section{Conclusion}
\label{con}
We have presented a new radio continuum study of \ac{LMC} \ac{SNR} \snr. We hypothesise the following:
%This \ac{SNR} shows:
\begin{itemize}
%\item The remnant's asymmetrical radio brightness morphology may be explained by a Type\,Ia single degenerate progenitor involving a white dwarf accreting material from a companion star which typically leaves behind interacting circumstellar material. 

%with a diameter of $\sim$8\,pc.(you don't need to tell them the diameter again, it has nothing to do with what you are saying here)

%Notes:for everything in this list, let's report data only related to what we want to say about the remnant as above, and for example why do we think it is a young remnant? Perhaps we tie it to spectral index.

\item \snr's polarisation is similar to younger remnants (order $10^2$ years) N\,103B in the \ac{LMC}, and, Kepler and G1.9+0.3 in the \ac{MW} Galaxy. Our remnant has an average fractional polarisation of $5\pm1$\% and $6\pm1$\% at 5500 and 9000\,MHz, respectively.

\item \snr\ has a spectral index of $\bm{-0.62\pm0.02}$, similar to those of other younger remnants including Kepler and SN\,1006.
\item We find estimates of the magnetic field strength of this remnant based on our data to be on the order of between 10 and 100\,$\mu$G, again similar to younger remnants including Kepler and Tycho. 

\item Based on \snr's position on our $\Sigma$-D track and its age as a young remnant, we suggest it may be at the end of its free expansion phase, entering the Sedov phase of evolution.
%\item Average \ac{RM} of $-139\pm86$\,rad\,m$^{-2}$.
%\item Magnetic field strength of 12.5\,$\mu$G, and  equipartition field of 71.7\,$\mu$G with an estimated minimum explosion energy of E$_{\rm min}$~=~2.6$\times10^{48}$\,erg.
\item The possible existence of a \HI\ cloud towards the south-east of the remnant and a suspected wind-bubble cavity in this region. 
%could possibly indicate the presence of an accreting white dwarf prior to the detonation of \snr, similar to that proposed for Tycho and N103B.
\end{itemize}

%This has been stated above, leave it out here: Based on the above findings, we favour a \ac{SN} Type\,Ia \acl{SD} scenario for \snr.% \Denis{ok if remarks last paragrah of section 3 are addressed}

\section*{Appendix: Velocity channel distributions of \HI\ }
\label{appendix}
\begin{figure*}[ht!]
\begin{center}
\includegraphics[width=\textwidth,clip]{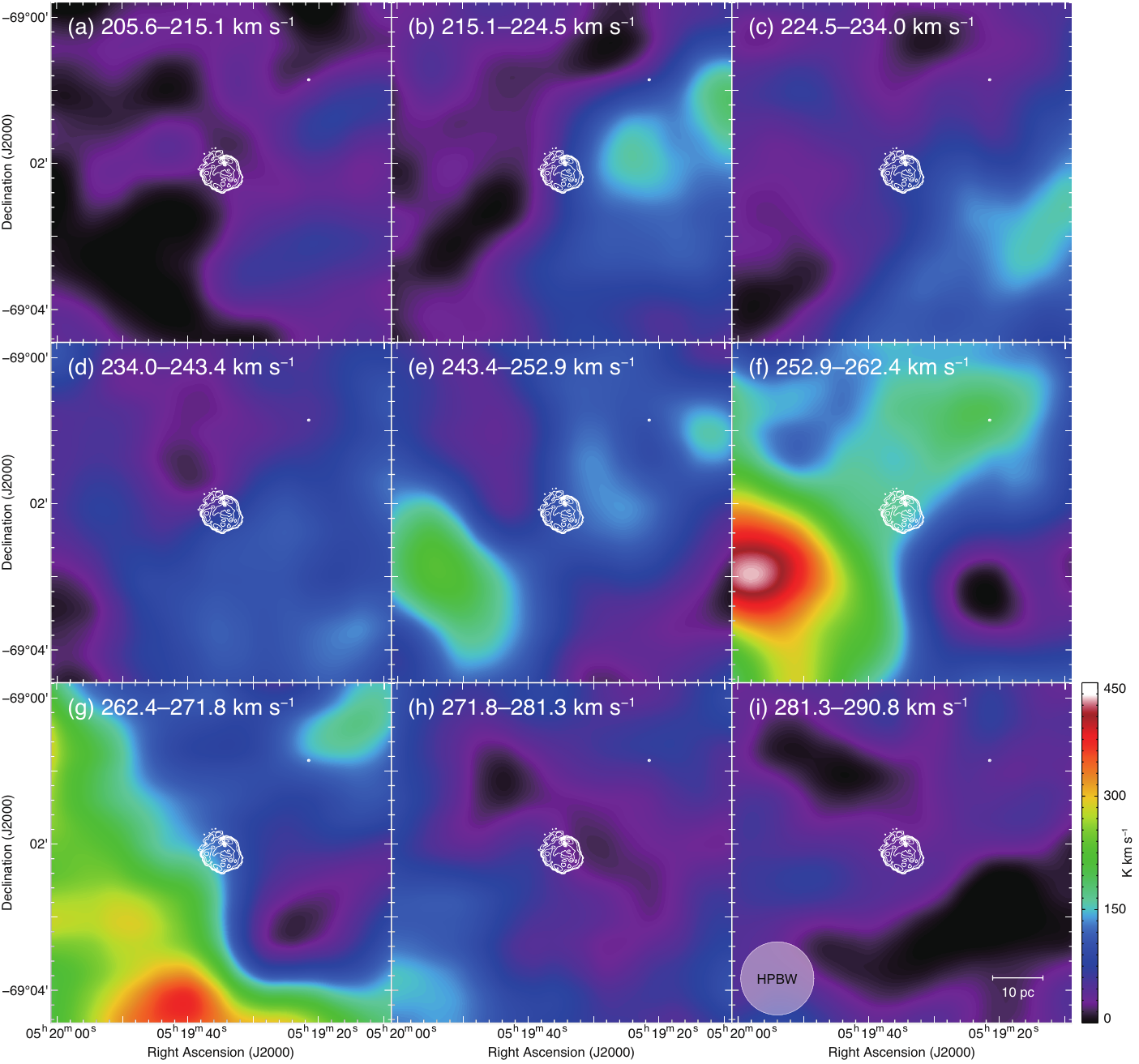}
\caption{Velocity channel distributions of \HI\ toward \snr. The superposed contours are the same as those shown in Figure~\ref{Fig1}. Each panel shows \HI\ distributions every 9.4\,km\,s$^{-1}$ in a velocity range from 205.6 to 290.8\,km\,s$^{-1}$.}
\label{fig_appendix}
\end{center}
\end{figure*}

Figure~\ref{fig_appendix} shows the velocity channel distributions of \HI\ toward \snr. We found \HI\ emission only in the velocity range from $\sim$220 to $\sim$270\,km\,s$^{-1}$. Although we could not confirm the cloud association with the \ac{SNR} due to the modest angular resolution of the \HI\ data, we can infer that the \ac{SNR} exploded in a relatively low-density gas environment. Further \HI\ observations with high-angular resolution are needed to investigate whether the \HI\ clouds are physically associated with the \ac{SNR}.

\section*{Acknowledgment}
The Australian Compact Array is part of the Australian Telescope which is funded by the Commonwealth of Australia for operation as National Facility managed by \ac{CSIRO}. This paper includes archived data obtained through the \ac{ATOA} (\url{http://atoa.atnf.csiro.au}). We used the \textsc{karma} and \textsc{miriad} software packages developed by the \ac{ATNF}.
MDF and GR acknowledge ARC funding through grant DP200100784.
D.U. acknowledges the Ministry of Education, Science and Technological Development of the Republic of Serbia through contract No. 451-03-68/2022-14/200104, and for the support through the joint project of the Serbian Academy of Sciences and Arts and Bulgarian Academy of Sciences on the detection of extragalactic \acp{SNR} and \HII\ regions.
H.S. was also supported by JSPS KAKENHI grant Nos. 19K14758, 21H01136, and 24H00246.

%\paragraph{Funding Statement}

%This research was supported by grants from the <funder-name> <doi> (<award ID>); <funder-name> <doi> (<award ID>).

%\paragraph{Competing Interests}

%A statement about any financial, professional, contractual or personal relationships or situations that could be perceived to impact the presentation of the work --- or `None' if none exist.

%\paragraph{Data Availability Statement}

%A statement about how to access data, code and other materials allowing users to understand, verify and replicate findings --- e.g. Replication data and code can be found in Harvard Dataverse: \verb+\url{https://doi.org/link}+.

\bibliography{references}

%\appendix
%\input{appendices}

\end{document}